# High-Throughput Imaging of Degradation-Inducing Microscopic Impurities in Perovskite Solar Cells

Sofiia Kosar[1,‡*], Anil R. Pininti[1], Vladyslav Hnapovskyi[1], José P. Jurado[1], Subhashri Mannar[1], Lorenzo Mardegan[1], Anand S. Subbiah[1], Frédéric Laquai[1,2], Stefaan De Wolf[1*]

[1] *Center for Renewable Energy and Storage Technologies (CREST), Physical Science and Engineering Division (PSE), King Abdullah University of Science and Technology (KAUST), Thuwal 23955-6900, Saudi Arabia*

[2] *Chair of Physical Chemistry and Spectroscopy of Energy Materials (SPECTRE), Department of Chemistry, Ludwig-Maximilians-University (LMU) Munich, Butenandtstraße 5-13, Munich 81377, Germany*

[‡] *Present address: Department of Chemistry and Center for Nanoscience (CeNS), Ludwig-Maximilians-University (LMU) Munich, Butenandtstraße 5-13, Munich 81377, Germany*

[*]Email: sofiia.kosar@kaust.edu.sa, stefaan.dewolf@kaust.edu.sa

**Abstract**

The scalable fabrication of high-quality, large-area perovskite thin films is hindered by microscopic inhomogeneities, particularly residual compositional impurities formed during processing. Identifying these impurities, understanding their impact on device operation, and enabling their rapid detection are essential for upscaling perovskite solar cells (PSCs). Here, nano-Fourier transform infrared spectroscopy combined with high-resolution optical and scanning probe techniques was used to identify detrimental impurities in wide-bandgap perovskite films relevant to tandem solar cells. Photo-stress experiments revealed that degradation of the perovskite layer initiates at impurity-perovskite interfaces, demonstrating that impurities act as failure nucleation sites. Leveraging these insights, a rapid, non-invasive, and high-throughput framework based on high-resolution reflected light microscopy and machine learning-supported image analysis was developed, enabling detection and quantification of harmful impurities in as-prepared films within seconds. Films with higher impurity density show accelerated early degradation, establishing this parameter as an early-warning metric for stability screening. Extending this framework to degradation tracking further reveals coupled photo- and thermo-chemical contributions to impurity-mediated instability. Overall, this work establishes a practical chemically-validated diagnostic imaging framework for rapid pre-screening of perovskite films to enable stable and scalable PSCs.

**Keywords:** perovskite, correlative microscopy, compositional impurities, high-throughput imaging.

## Main

Perovskite solar cells (PSCs) have demonstrated extraordinary potential for photovoltaic (PV) applications, with power conversion efficiencies (PCEs) exceeding 27 % for single-junction devices, and 34 % for perovskite/silicon tandem solar cells[1]. However, their widespread deployment remains hindered by challenges associated with long-term operational stability[2-4] and reproducible upscaling to industrially relevant areas[5-7]. As the field transitions from laboratory-scale spin coating to scalable deposition methods (e.g., blade-coating, slot-die coating, or vacuum processing)[8], the need for rapid, predictive quality control of perovskite thin films becomes increasingly urgent.

Addressing these challenges requires not only improvements in intrinsic material robustness but also rigorous control over microscopic film quality. Processing-induced inhomogeneities – including pinholes[9,10], wrinkles[11-13] and compositional variations[14-16] – can compromise device performance and obscure stable and scalable PSC fabrication[17]. In particular, remnant photo-inactive impurity phases persisting on the surfaces of the perovskite layers[18-21], can locally perturb opto-electronic properties, act as non-radiative recombination centers[19,22], and serve as initiation sites for photo-degradation and catastrophic failure[20,23-25]. Their impact is especially critical for mixed-cation mixed halide wide-band gap compositions - central for tandem applications [26] - where complex compositional engineering and elevated annealing temperatures increase the likelihood of forming residual photo-inactive phases[27]. Despite this, systematic integration of impurity chemistry into rapid stability screening workflows remains unresolved.

High-resolution techniques capable of capturing microscopic impurities such as scanning probe microscopies and nano-spectroscopies provide unparalleled insights into local morphology, structure and chemistry[28,29] but are inherently slow (i.e., often requiring hours to scan micrometer-sized regions) and incompatible with high-throughput in-line diagnostics that can only afford seconds of characterization time per sample[30]. Conversely, camera-based imaging approaches - highly-promising for in-line diagnostics of PSCs[31-36] - despite faster operation and ability to capture large-areas, lack spatial resolution and chemical specificity needed to identify microscopic stability-relevant impurities and generally remain noncalibrated for stability predictions. This creates a stability screening gap: microscopic impurity phases may escape detection during fabrication yet later act as deterministic degradation seeds under operational stress. Establishing a chemically-validated and non-invasive high-throughput screening framework capable of identifying and quantifying stability-relevant impurities prior to device completion is therefore essential.

Here, we address these challenges by investigating impurity-driven degradation in blade-coated ~1.68 eV CsFA-rich mixed-halide perovskite films relevant for perovskite/silicon tandem applications. Using a suite of correlative scanning-probe- and optical- microscopies, we identify remnant $PbI_2$ and 4H hexagonal perovskite polytype impurities on the surfaces of our films and elucidate their roles as degradation triggers under photo-stress. We then translate our fundamental insights to develop a rapid, all-optical reflected light microscopy (RLM) framework combined with convolutional neural network (CNN) image analysis to enable high-throughput, non-destructive and quantitative screening of compositional impurities in as-prepared films within seconds. Films with dense impurity distribution exhibit accelerated early degradation, establishing impurity density as an early-warning metric for stability screening prior to device fabrication. By extending this framework to degradation tracking, we reveal that impurity-mediated instabilities arise from coupled photo- and thermo-chemical processes. Together, these results provide both mechanistic insights into microscopic degradation pathways and establish a practical diagnostic toolkit for accelerating the upscaling of perovskite PV technologies, including tandems.

**Mapping compositional impurities in blade-coated wide bandgap perovskite films**

We blade-coated $Cs_{0.22}FA_{0.63}MA_{0.15}Pb(I_{0.83}Br_{0.14}Cl_{0.03})_3$ films[37] on glass/ITO/SAM (ITO is indium tin oxide, SAM is a self-assembled monolayer, Me-4PACz in our case[38]) substrates and used nano Fourier-transform infrared spectroscopy (nano-FTIR) based on a scattering-type scanning near-field optical microscope (s-SNOM) to gain nanoscale insights into the local morphology and chemical composition of the perovskite films in device-relevant layer stacks (Fig. 1a, Supplementary Note 1)[18]. The morphology maps collected across multiple regions and samples showed that the films were mostly composed of pristine grains that were compact and varied in size from 500 nm to 1 µm (Fig. 1b, Supplementary Fig. 1a). Apart from pristine grains, we frequently observed hexagonally-shaped clusters with dimensions up to several micrometers (Fig. 1c, green box, Supplementary Fig. 1b), and isolated grain-like features that were smaller (300 - 500 nm) and exhibited altered morphologies distinct from pristine grains (Fig. 1d, blue box, Supplementary Fig.

1c). To gain insight into the origin of the detected impurity features, we collected local chemical information. White-light infrared (IR) maps collected simultaneously alongside topography maps, provide qualitative first-glance information about the chemical activity of materials, where weak IR scattering signals are indicative of strong IR light absorption, inherent to materials containing organic components. Fig. 1e-g and Supplementary Fig. 1d-f show that pristine grains were associated with uniform and strong IR absorption, seen from the weak IR scattering, while hexagonally-shaped clusters and isolated impurities exhibited a complete lack of or only a weak IR absorption, respectively, seen from their stronger IR scattering signals. The nano-FTIR spectra further showed that IR absorption of the pristine grains originated from the vibrational activity of the C=N bond of the formamidinium ($FA^+$) cation, which showed characteristic strong vibrational resonances at ~1714 $cm^{-1}$ (Fig. 1h, Supplementary Fig. 1g). This resonance was completely absent for hexagonal clustered impurities (Fig. 1i, Supplementary Fig. 1h), suggesting their inorganic nature, and weak for the isolated impurities (Fig. 1j, Supplementary Fig. 1i), emphasizing their different chemical nature compared to pristine grains. The $MA^+$ cation, due to its small fraction, was not detected at the nanoscale (see Supplementary Fig. 2 for the whole spectrum). Considering the FA-rich nature of our films, we focus our discussion on the presence of $FA^+$ as an indicator of a pristine perovskite phase.

To further understand the nature of observed impurities, we performed X-ray diffraction (XRD) measurements (Fig. 1k, Supplementary Fig. 3). Based on the detected XRD patterns, we attributed the observed impurities to residual $PbI_2$ (2Θ = 12.7°) and 4H hexagonal perovskite polytype phase (2Θ = 11.3°) that are common by-products in mixed halide perovskite thin films[39,40] and are known to remain on the surfaces of perovskite films as nanoscale inclusions[18,19,23]. Interestingly, in our ~1.68 eV blade-coated films, the inclusions of $PbI_2$ formed large crystalline clusters, leading to appreciable characteristic XRD peak, which we hypothesize to stem from the high annealing temperatures (150 °C) required to achieve the cubic phase for CsFA-rich perovskite compositions. The inclusions of the 4H polytype phase are barely detectable in conventional bulk XRD, consistent with their sparse distribution on the surface and nanoscale sizes yet can be readily detected with table-top surface-sensitive nano-FTIR. Both impurities were found to persist upon A-site cation and additive modifications (see Supplementary Fig. 4-6 and Supplementary Note 2).

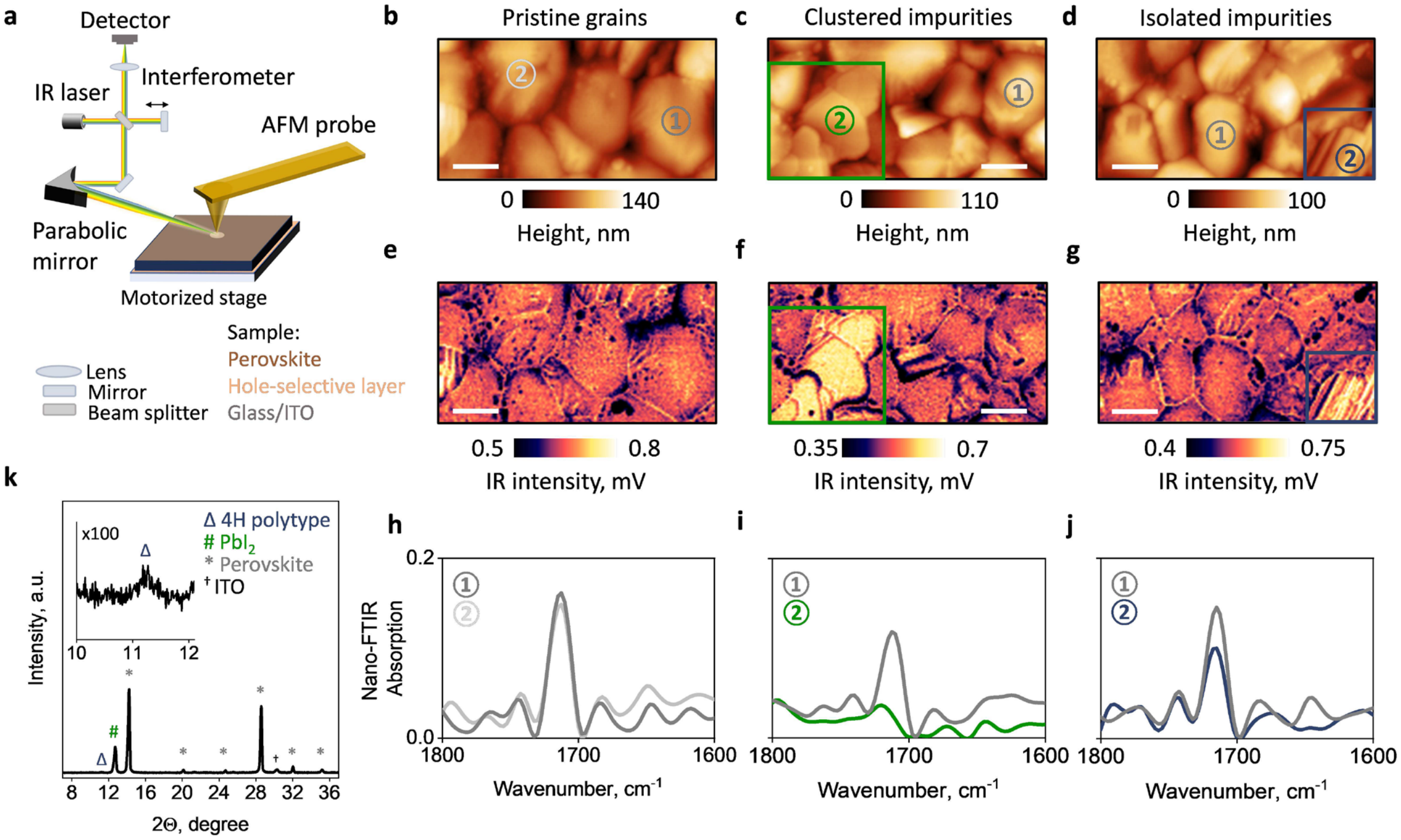

**Figure 1. Identification of nanoscale compositional impurities.** (a) Schematic of nano-FTIR setup employed to characterize local morphology and composition of blade-coated perovskite films. AFM maps of regions with (b) pristine grains, (c) hexagonally-shaped clustered impurities (marked with a green box), and (d) with an isolated impurity (marked with a blue box). (e-g) White-light IR maps corresponding to the regions in (b-d), showing strong and uniform IR absorption by the pristine grains (seen as a weak IR scattering signal), and weak IR absorption by the clustered and isolated impurities (seen as a strong IR scattering signal). (h-j) Nano-FTIR spectra collected from points of interest as indicated in AFM images in (b-d). Pristine perovskite grains are marked with grey, hexagonal clustered impurities with green, and an isolated impurity with blue. Scale bars are 500 nm. (k) XRD of a blade-coated perovskite film showing the presence of a dominant cubic perovskite phase (marked with "*"), and inclusions of remnant $PbI_2$ (marked with "#") and 4H hexagonal perovskite polytype phase (marked with "Δ"). Magnified region highlighting 4H polytype phase is shown in the inset.

### Impact of compositional impurities on performance and stability of PSCs

Having identified remnant microscopic compositional impurities, we now study their specific impacts on the opto-electronic performance and stability of WBG PSCs. The lower contact potential differences (CPDs) associated with compositional impurities, mapped with Kelvin probe force microscopy (KPFM) (Fig. 2a, b), indicate larger work functions resulting in upward band bending at the perovskite-impurity interface, which together with the WBG nature of impurities may lead to an energetic barrier for electron transport once an electron transport layer is introduced. High-resolution confocal photoluminescence (PL) microscopy revealed an inhomogeneous PL response with weakened intensity in the areas associated with impurities (Fig. 2c, see Supplementary Fig. 7 for PL spectrum), attributed to their photo-inactive and likely defect-rich nature[19,22]. Collectively, these impurity-induced variations in surface potential and PL are expected to impede the operating voltage of PSCs (see Supplementary Fig. 8 for device characteristics) and thus must be avoided in industrially-relevant devices.

The impact of the remnant compositional impurities on stability of WBG PSCs was investigated via *ex-situ* morphological and compositional mapping, capturing the progression of accelerated photo-degradation. We exposed the films to white light with a 3-Sun equivalent intensity in an inert ($N_2$) atmosphere and mapped them at several locations along the diverging incident beam, thus gaining insight into the progression of degradation. We observed that nanoscale granular structures developed at the sites of initial compositional impurities and eventually seized the entire impurity clusters, while initially-pristine grains remained intact (Fig. 3d). Further, this impurity-initiated degradation, propagated to the interface with pristine grains (Fig. 3e), causing irreversible morphological changes of the pristine material, manifested as grain decomposition. Chemically, through local nano-FTIR spectroscopy, we found that the decomposed areas exhibited complete depletion of $FA^+$, seen from the absence of C=N resonances, while the grains that remained morphologically pristine in the intermediate phase still showed C=N resonances and thus preserved the perovskite structure (Fig. 2f). To gain chemical insight into degradation by-products, we collected confocal Raman spectra following the prolonged photo-stress with additional heating at 85 °C (Fig. 2g). We observed that upon complete morphological decomposition, the films mostly showed signals of $PbI_2$ with characteristic Raman peaks at 96 $cm^{-1}$ and 115 $cm^{-1}$ [41,42], and an intermediate, likely Cs-rich, phase with broad peaks at 45 $cm^{-1}$ and 70 $cm^{-1}$ [43]. Some remaining pristine perovskite material with a characteristic peak at 29 $cm^{-1}$ was also observed upon complete degradation[44].

Our microscopic findings reveal impurity-induced degradation consistent with earlier reports for 1.54 - 1.63 eV spin-coated and vacuum-evaporated perovskite films[23]. Our own observations for ~1.68 eV blade-coated films highlight the detrimental roles of remnant compositional impurities regardless of the fabrication method of perovskite films and show that especially for WBG CsFA-rich compositions, requiring high annealing temperatures, the control of impurities becomes critical, as they tend to cluster on the film surface. This highlights the critical importance of assessing impurity distribution to ensure device stability when developing new compositions and adapting them for scalable and tandem applications.

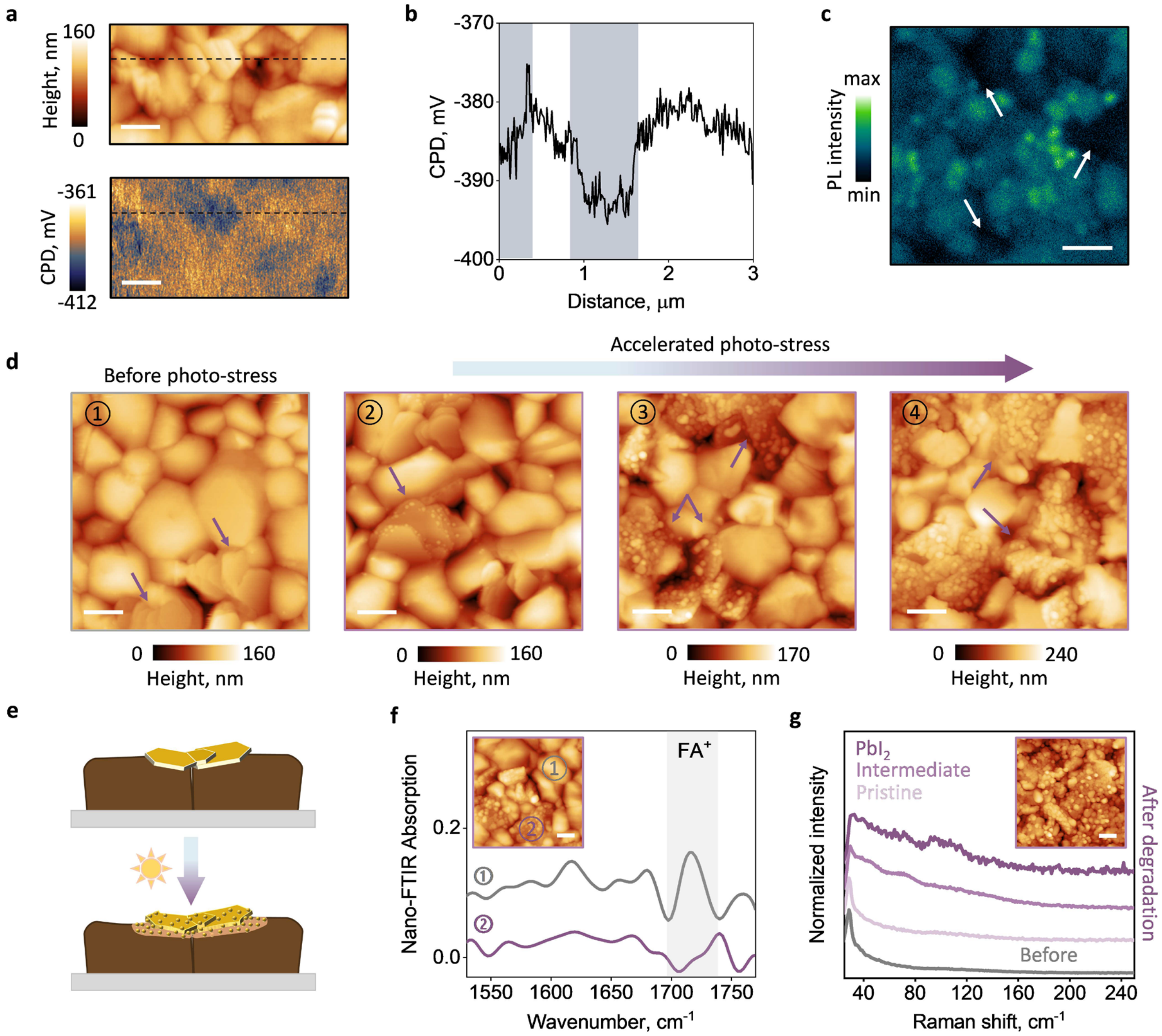

**Figure 2. Impact of microscopic compositional impurities on performance and stability of PSCs.** (a) AFM (top) KPFM (bottom) maps of perovskite thin film. (b) CPD line profile capturing compositional impurities and pristine grains, showing that impurities are associated with larger CPD values (highlighted with grey). Scale bars are 500 nm. (c) Confocal PL map of perovskite thin film showing uniform PL within the pristine grains and weakened PL intensity associated with impurities (marked with white arrows). Scaler bar is 2 µm. (d) AFM images featuring microscopic progression of degradation. Before illumination (①), the film is composed of pristine grains and impurities (marked with purple arrows). As illumination progresses, nanoscale speckles develop within the impurities (②) and further strike the entire impurity features (③), while pristine grains keep their initial compact morphology. With further illumination, the decomposition propagates into the pristine grains (④), causing their irreversible fragmentation. Scale bars are 500 nm. (e) Schematic illustration of the degradation process via fragmentation upon illumination. (f) Nano-FTIR spectra of a compact grain (grey) and decomposed area (purple), as indicated in the AFM image in the inset, showing the absence of $FA^+$ in the degraded region. (g) Confocal Raman spectra of perovskite film before and after complete degradation with an AFM image of the degraded film in the inset. Scale bars for inset images are 500 nm.

**High-throughput imaging of compositional impurities and CNN analysis**

Having identified microscopic compositional impurities that promote photo-degradation, we stress the importance of controlling their presence already at the stage of perovskite film fabrication, before processing subsequent device layers. This step is critical to evaluate the film quality and predict its susceptibility to impurity-induced degradation. This requires the ability to detect and quantify the compositional impurities rapidly and non-invasively in a high-throughput fashion. This, however, is challenging for advanced mapping techniques including scanning probe microscopies, known for their generally slow mapping times and limited compatibility with in-line high-throughput characterization required for industrial applications.

Given the unique chemical nature of compositional impurities, their refractive indices are expected to differ from those of pristine perovskite grains, which offers opportunities for identifying them using optical imaging techniques with rapid frame acquisition. As such, we first mapped our films with nano-FTIR and established the presence of compositional impurities seen as distinct features in morphology and white-light IR scattering maps (Fig. 3a, b). From local nano-FTIR spectra, we attributed the observed impurities, marked "1" and "3" in Fig. 3a and c, to residual 4H perovskite polytype and $PbI_2$ phases, respectively, based on their weak and absent C=N resonances, respectively. Pristine grains marked "2", "4", and "5" in Fig. 3a and c exhibited strong C=N resonances, indicating their FA-rich nature. We then imaged the same region with high-magnification RLM and found that optical contrast was inhomogeneous, showing microscopic areas with enhanced reflectance (Fig. 3d). Upon superimposing the RLM images and white-light IR maps (Fig. 3e), we revealed that microscopic compositional impurities correlated strongly with areas showing enhanced reflectance, giving rise to distinct contrast, consistent with the distinctive refractive indices of compositional impurities. Importantly, this contrast is preserved in grayscale (Fig. 3f), indicating that impurity detection is also feasible with monochrome cameras. Additional experiments on multiple areas and films confirmed that optically compositional impurities manifest themselves as features with enhanced reflectance (see Supplementary Fig. 9 and 10 for additional correlative nano-FTIR/RLM and AFM/RLM analysis). As such, rapid and non-invasive RLM can be employed for high-throughput impurity identification and quality assessment of as-prepared perovskite thin films.

We further leveraged machine-learning based methods to develop image analysis for impurity identification that can be integrated into the perovskite fabrication lines along with RLM imaging. Our method is based on a supervised convolutional neural network (CNN)[45,46] with U-Net structure[47] that was trained to read RLM images of perovskite films, analyze their contrast, and output masks with identified compositional impurities. To prepare a training set, we segmented RLM images from the microscope into a sub-set of smaller images (256 × 256 pixels) (Fig. 3g and Supplementary Fig. 11a). To annotate impurities in each RLM image - required for efficient training - we prepared labels by thresholding the greyscale images until only compositional impurities remained highlighted (Fig. 3g and Supplementary Fig. 11a). The CNN was built in Python using Tensorflow[48] and Keras[49] and contained three encoder and decoder layers (Supplementary Fig. 11b). To evaluate the accuracy of the model, we used binary cross-entropy (BCE) loss function and Dice-Sørensen coefficient (DSC) which reached values of ~ 0.06 and ~ 0.91, respectively (Fig. 3h). Both the low BCE loss function and high DSC are signatures of efficient training. Fig. 3i shows an RLM image of a perovskite film with multiple microscopic impurities on the surface, previously unseen by the model (see Supplementary Fig. 11c-e for images from the test set). The trained model promptly identified compositional impurities and produced a corresponding impurity mask (Fig. 3j), which accurately follows the impurity distribution (Fig. 3k). Supplementary Fig. 12-14 and Supplementary Note 3 discuss nano-FTIR/AFM/RLM correlations and analysis of additional sample types, highlighting the robustness of model predictions. With complementary post-analysis, we further estimated the total impurity coverage ($C_I$), impurity sizes ($S_I$), and nearest-neighbor distances ($NND_{I-I}$) among impurities, which we further implemented for quantitative analysis (Fig. 3l).

We used the developed framework for quantitative analysis of impurity distribution upon bulk and surface modifications in blade-coated perovskite films in device-relevant stacks (Supplementary Fig. 15, 16 and Supplementary Note 4). We found that while removal of bulk additives leads to smaller impurities, their distribution becomes denser, increasing the abundance of future degradation sites. The post-treatment of films with propane-1,3-diammonium iodide (PDAI), while decreased the impurity sizes, was found to only marginally impact $NND_{I-I}$ and was overall insufficient to mitigate surface impurities in contrast to several reports [50-52]. The developed RLM framework thus offers rapid screening of as-prepared films with image acquisition and analysis time of several seconds, opening a clear pathway for high-throughput non-destructive diagnostics of perovskite films, suitable for both in-lab and industrial applications.

To further demonstrate the effect of impurity-induced degradation on device stability, we fabricated single-junction PSCs based on films with and without bulk modification (Fig. 3m, n). The device based on the film without bulk modification, exhibiting dense impurity distribution, $NND_{I-I}$ of 729 nm, showed early signs of degradation within the first hour of operation (Fig. 3o, Supplementary Fig. 17). The degradation primarily originated from the rapid drop in short-circuit current density ($J_{SC}$) and open-circuit voltage ($V_{OC}$), consistent with the decomposition of perovskite facilitated by the high density of impurities (Fig. 3p). In contrast, the device based on the film with bulk modifications and moderate impurity density, $NND_{I-I}$ of 900 nm, exhibited significantly improved short-term operational stability, indicating a much slower progression of degradation. These results directly link the optically-identified impurity distribution to device-level degradation behavior, further validating our method as a predictive diagnostic tool for perovskite film quality, and establishing impurity density ($NND_{I-I}$) as an early-warning metric for stability screening prior to device fabrication.

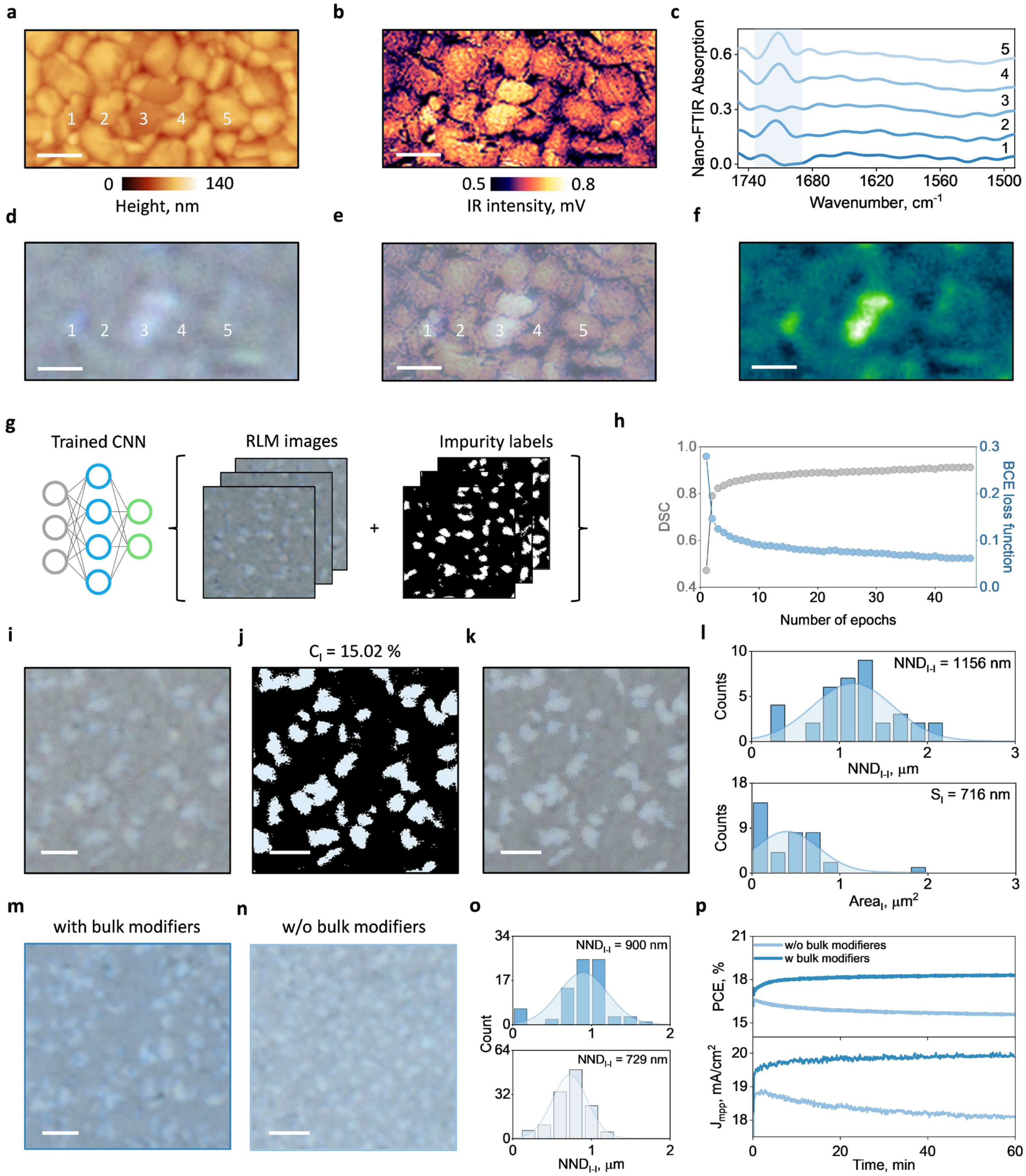

**Figure 3. High-throughput identification of compositional impurities**. AFM (a) and white-light IR (b) maps of perovskite thin film. (c) Nano-FTIR spectra collected from the features marked in (a), representing compositional impurities ("1", "3") and pristine grains ("2", "4", and "5"). (d) RLM image of the region shown in (a) and (b) with features of interest marked. (e) Superimposed RLM image and white-light IR map from the same region as in (a) and (b) showing that features with enhanced optical reflectance correlate well with compositional impurities. (f) RLM image from the same region shown in greyscale. (g) U-Net model training for impurity identification. (h) BCE loss function and DSC indicating effectiveness of the training. (i) RLM image of a blade-coated perovskite film previously unseen by the U-Net model. (j)

Impurity mask for (i) predicted by the model with total impurity coverage ($C_I$). (k) Superimposed RLM image and impurity mask showing high accuracy of the model prediction. (l) Nearest neighbor distance ($NND_{I-I}$) and impurity size ($S_I$) distributions calculated for (i). (m) RLM image of blade-coated film with bulk modifications leading to moderate impurity density. (n) RLM image of a blade-coated film without bulk modifications leading to dense impurity distribution. (o) $NND_{I-I}$ for RLM image in (m) shown with dark blue (top), and for (n) – shown with light blue (bottom). (p) Device stability data showing rapid degradation of the device based on perovskite film without bulk modifiers that is associated with dense impurity distribution. Scale bars for (a-f) are 1 µm, for (i-k) and (m, n) - 2 µm.

**Tracing impurity-induced degradation with high-throughput imaging**

Having established compositional impurities as degradation initiation sites and quantified their spatial density using a rapid RLM framework, we next extend this approach to monitor impurity-mediated degradation under device-relevant stress conditions. As we demonstrated with nano-FTIR analysis, impurity-induced degradation of perovskite films originates at impurity clusters and propagates as alteration of initially-compact grain morphology at the impurity-perovskite junctions, accompanied by the loss of $FA^+$. Such compositional changes of perovskite grains are expected to alter the local refractive indices, and as such, the degradation can be also traced using RLM. Fig. 4a shows an RLM image of a device-relevant layer stack terminated with a perovskite film after photo-stress with white light with a 3-Sun equivalent intensity. The illumination was conducted from the film side, as described in Fig. 3d, and the image was collected at the intermediate stage (marked with ③ in Fig. 3d). Local darkening of RLM contrast was observed selectively at the junctions of impurity clusters and adjacent grains, which we attribute to enhanced scattering from the decomposing perovskite material, consistent with nano-FTIR observations. The pristine grains located away from the impurity clusters remained optically unchanged (see magnified RLM image in Fig. 4a). The RLM images corresponding to all the stages ①-④ of photo-stress are shown in Supplementary Fig. 18.

To quantitatively map the degraded regions, we adapted our U-Net model to identify areas associated with local darkening of RLM contrast (Supplementary Fig. 19). Subsequent superposition of impurity- and degradation- masks produced by respective U-Net models, revealed that degraded regions were consistently located adjacent to the pre-existing impurities (Fig. 4b, Supplementary Fig. 20). The extracted impurity-to-degradation nearest neighbor distances ($NND_{I-D}$) showed the mean value of 869 nm, smaller than the $NND_{I-I}$ (967 nm) and comparable to the average grain size of ~ 800 nm (Fig. 4c). This quantitative analysis confirms that degradation was not randomly distributed but was spatially anchored to initial impurity sites.

To assess the degradation behavior under conditions relevant for single-junction PSCs, we conducted photo-stress experiments with illumination incident from the glass side. Interestingly, we observed similar behavior as in the case of the film-side illumination. As shown in Fig. 4d, following 2 hours of photo-stress, local darkening developed preferentially at the impurity-perovskite junctions, and, to a lesser extent, along some of the grain boundaries, while large regions with pristine grains preserved their initial morphology and contrast (see Supplementary Fig. 21 and Supplementary Note 5 for correlative image analysis for 1-3 hours of photo-stress). The predicted impurity-degradation map confirmed their close spatial correlation with mean $NND_{I-D}$ of about 833 nm compared to an $NND_{I-I}$ of 1091 nm (Fig. 4e, f). Even after extended illumination (7 hours), pristine grains remained clearly discernible, whereas degraded regions remained localized around the initial impurity clusters (Supplementary Fig. 22 and Supplementary Note 5).

We then subjected our sub-device stacks to heating at 85 °C in the absence of illumination. After only 1 h of heating, RLM images revealed the emergence of localized darkened regions at impurity-perovskite junctions (Fig. 4g and h, Supplementary Fig. 23 and Supplementary Note 5), closely resembling the degradation patterns observed under illumination. We note that pristine grains remained intact, emphasizing that impurities act as preferential activation sites for degradation even in the absence of photo-excitation. The impurity-degradation distribution indicated similar behavior as in the case of photo-stress with $NND_{I-D}$ of 862 nm and $NND_{I-I}$ of 1165 nm. Consistent behavior was observed upon varying the annealing duration of our blade-coated films beyond 20 minutes at 150 °C. Supplementary Fig. 24 shows progressive appearance of microscopic optically-dark regions, concomitant with emergence of δ-$CsPbI_3$ peak (at 2Θ = 10°) in XRD.

We also note that upon shelf storage in inert $N_2$ environment, our films remained stable, showing no changes following 10 months of storage (Supplementary Fig. 25). All the microscopic optical changes were observed exclusively as a result of light or heat stress. Our methodology thus paves the way not only for impurity tracking, but also for rapid and non-invasive monitoring of degradation. Importantly, RLM-based impurity and degradation imaging remains effective not only for device-relevant layer stacks but also for full semi-transparent devices, where impurities and perovskite layers can be imaged through the transparent top electrodes (Supplementary Fig. 26).

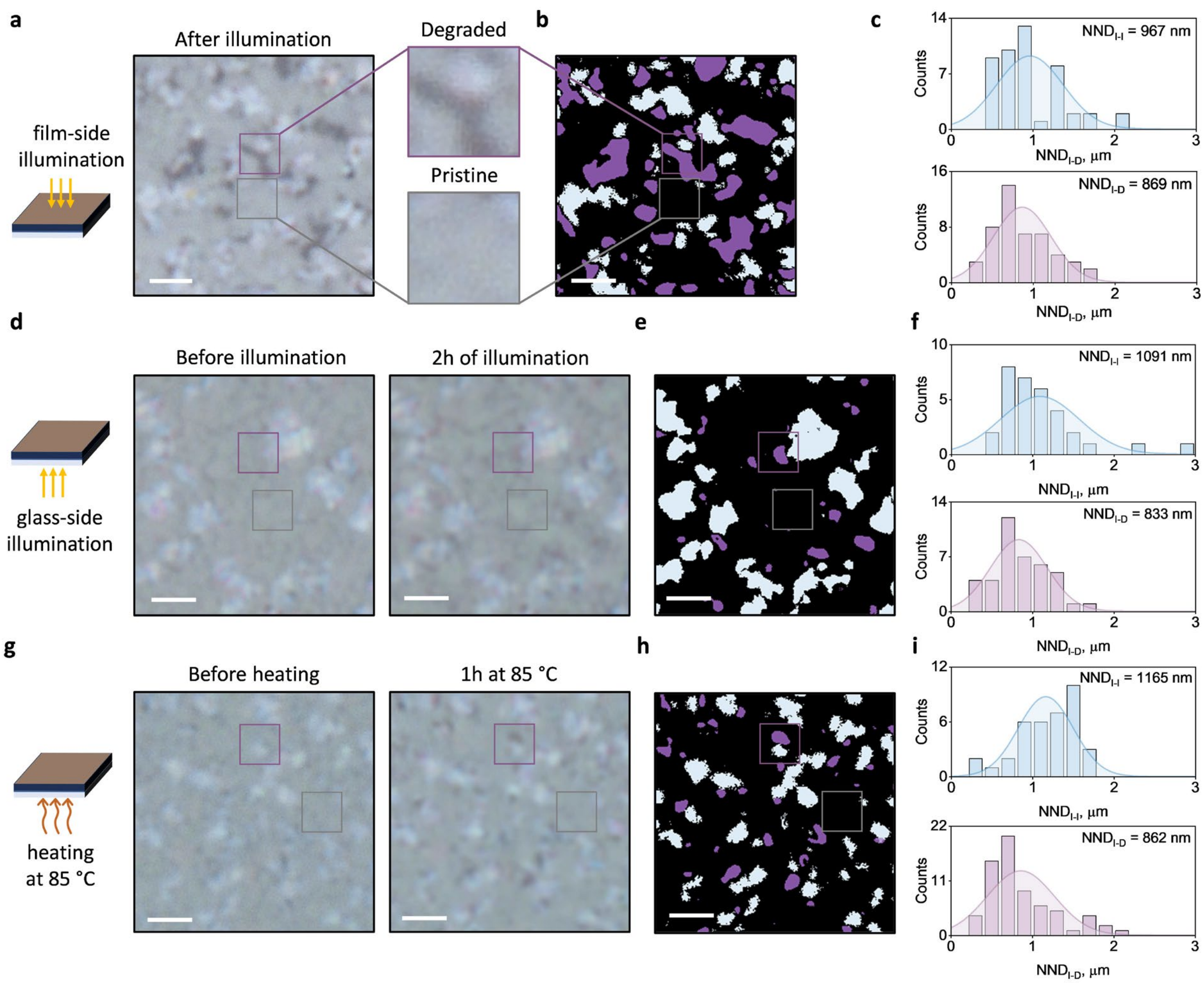

**Figure 4. Optical monitoring of impurity-induced degradation.** (a) RLM image of a blade-coated perovskite film at intermediate stage of illumination from the film side. Purple and grey boxes highlight degraded and pristine areas, respectively, that are shown as magnified images on the right. (b) Impurity-degradation map for (a) obtained by superimposing impurity (light blue) and degradation (purple) maps produced by respective U-Net models. (c) Nearest neighbor distance distributions among impurities ($NND_{I-I}$) and impurity-to-degradation nearest neighbor distance distribution ($NND_{I-D}$) for (a), highlighting occurrence of degradation at the impurity-perovskite junctions. (d) RLM images of perovskite film before and after 2 hours of illumination from the glass side. (e) Impurity-degradation map for (d). (f) $NND_{I-I}$ and $NND_{I-D}$ distributions for (d). (g) RLM images of perovskite film before and after 1 hour of thermal stress at 85 °C. (h) Impurity-degradation map for (g). (i) $NND_{I-I}$ and $NND_{I-D}$ distributions for (g). Scale bars are 2 µm.

We summarize the possible impurity-facilitated degradation mechanisms as follows. Under top illumination, where the incident light directly excites both the perovskite layer and impurities, photo-chemical processes dominate[53,54]. Illumination enables the interaction of photo-excited carriers with defects in the 4H perovskite polytype impurities and promotes defect-assisted photolysis of $PbI_2$, leading to the formation of $Pb^0$ and molecular iodine[23]. These species locally destabilize perovskite, driving $FA^+$ effusion from the lattice and morphological decomposition at the impurity-perovskite junctions. For the glass-side illumination, photo-chemical processes are attenuated because the high-energy photons required to induce $PbI_2$ photolysis are absorbed by the glass substrate and the uppermost ~100 nm of the active layer. Although photo-excited carriers can still interact with the defective impurities, photo-chemistry alone cannot fully account for the observed degradation. Instead, heating plays a critical and often overlooked role: temperature rise under illumination activates thermo-chemical reactions, accelerates halide and cation migration, and enhances chemical reactivity at impurity-perovskite junctions, promoting $FA^+$ loss and grain decomposition. This interpretation is consistent with the fact that heating at 85 °C produces nearly identical microscopic degradation patterns to illumination. We thus conclude that impurity-driven instability arises from coupled photo- and thermo- chemical processes and heating effects cannot be neglected in assessing impurity-mediated degradation pathways. As such, remnant compositional impurities will inevitably contribute to long-term instability regardless of illumination geometry. Consequently, minimizing impurity formation at the stage of film fabrication is essential for achieving stable PSCs. The rapid RLM imaging coupled with CNN-based analysis introduced in this work thus offers a powerful and practical platform for early, non-invasive diagnostics of detrimental stability-limiting microscopic impurities (see Supplementary Note 6 for comparison with other imaging techniques for stability-relevant quality assessment).

**Discussion**

Our results have important implications for the field of perovskite PV. They underscore that achieving apparent quality in bulk characterization is insufficient to ensure stability, and that microscopic impurity control is essential, particularly for perovskite compositions processed at elevated temperatures. Specifically, through correlative scanning probe- and optical- microscopy, we demonstrated that remnant microscopic compositional impurities play a decisive role in governing the stability of PSCs by acting as active precursors to future device failure.

Importantly, we introduced a rapid and non-invasive, contactless and fully-optical approach for identifying detrimental impurities in as-prepared films. When combined with CNN-based image analysis, this framework enables quantitative assessment of impurity distribution with acquisition and analysis times of several seconds, offering a practical route toward early-stage pre-screening of perovskite films. This framework was also successfully extended to monitor degradation revealing that thermal effects under illumination - often overlooked - are as critical as photo-chemistry in impacting the impurity-driven degradation of PSCs.

Looking ahead, integrating RLM and machine learning-supported diagnostics directly into in-line and robotic coating platforms could significantly accelerate process optimization and microscopic quality control for industrial PSCs production (Supplementary Fig. 27). Furthermore, the application of this methodology for real-time monitoring of degradation can strengthen the stability testing approaches by bringing microscopic insight into the failure points. Beyond PSCs, the general approach of correlating nanoscale chemical identification with fast optical imaging may be extended to other emerging thin-film photovoltaic- and optoelectronic- materials, where microscopic heterogeneities limit the performance and long-term stability.

## Methods

### *Materials*

Full-area and patterned 1-inch$^2$ glass/ITO substrates (sheet resistance ≈ 15 Ω $sq^{-1}$) were supplied by Xin Yan Technology Ltd. Lead based precursors $PbI_2$ (99.999%), $PbBr_2$ (99.999%), $PbCl_2$ (99.999%), and $Pb(SCN)_2$ (99.5%), were obtained from Fisher Scientific. Cesium iodide (CsI, 99.999%), ethanol (EtOH, absolute, 99.5%), isopropyl alcohol (IPA, anhydrous, 99.5%), 1,3-dimethyl-2-imidazolidinone (DMI, 99.0%), and N,N-dimethylformamide (DMF, anhydrous, 99.8%) were purchased from Sigma-Aldrich. Organic halide salts, including methylammonium bromide (MABr), formamidinium iodide (FAI), methylammonium chloride (MACl), and propane-1,3-diammonium iodide (PDAI), were sourced from Greatcell Solar Materials. The self-assembled monolayer molecule 4-(3,6-dimethyl-9H-carbazol-9-yl)butyl phosphonic acid (Me-4PACz) was obtained from Luminescence Technology Corp. All chemicals were used as received without further purification.

### *Precursor solution preparation*

A control Me-4PACz solution (1 mM) was prepared by dissolving 0.331 mg of Me-4PACz in 1 mL of absolute ethanol. For triple-cation CsFAMA without additives formulation, a 1.8 M precursor solution with the composition $Cs_{0.22}FA_{0.63}MA_{0.15}Pb(I_{0.83}Br_{0.14}Cl_{0.03})_3$ was prepared. Appropriate stoichiometric amounts of CsI, FAI, MABr, MACl, $PbI_2$, $PbBr_2$, $PbCl_2$, and $Pb(SCN)_2$ were dissolved in a mixed solvent system consisting of 950 µL DMF and 50 µL DMI at room temperature. To promote controlled crystallization during coating, 10 mol% excess MACl and 1.5 mol% $Pb(SCN)_2$ (relative to $Pb^{2+}$) were incorporated as bulk additives. For triple-cation CsFAMA with additives formulation, the above stock solution was modified by introducing 10 µL each of additional additives, including phenethylammonium bromide (PEABr, 40 mg $mL^{-1}$ solution), potassium thiocyanate (KSCN, 15 mg $mL^{-1}$), urea (15 mg $mL^{-1}$), and potassium iodide (KI, 9 mg $mL^{-1}$). For double-cation Cs-free formulation, a 1.8 M precursor solution with composition $FA_{0.85}MA_{0.15}Pb(I_{0.85}Br_{0.15})_3$ was prepared by dissolving FAI, MABr, MACl, $PbI_2$, $PbBr_2$, and $Pb(SCN)_2$ in the same DMF/DMI solvent mixture. 10 mol% excess MACl and 1.5 mol% $Pb(SCN)_2$ were included to enhance film formation during blade-coating.

### *Perovskite film fabrication*

Patterned 1-inch$^2$ glass/ITO substrates were sequentially cleaned and treated with UV-ozone for 15 min before transfer into a nitrogen-filled glovebox. The hole-transport layer (Me-4PACz) was deposited by spin coating at 5000 rpm for 30 s and subsequently annealed at 100 °C for 10 min. After cooling to room temperature, the films were rinsed with ethanol under identical spin-coating conditions to remove excess material.

Perovskite layers were deposited by blade coating using 9.5 µL of precursor ink, with a blade-to-substrate gap of 100 µm and a coating speed of 18 mm $s^{-1}$. Immediately after deposition, the wet films were subjected to a nitrogen air-knife treatment to facilitate uniform solvent removal and crystallization. Thermal annealing was carried out in two steps at 75 °C for 30 min followed by 150 °C for 20 min, with annealing durations adjusted where specified. Where indicated, a surface passivation layer based on PDAI (0.2 mg $mL^{-1}$) was applied by spin coating and annealed at 100 °C for 10 min.

For CsFA perovskite formulation processed via hybrid method, pre-patterned glass/ITO substrates were exposed to a UV-$O_3$ lamp for 15 minutes. Subsequently, a 1 mg/mL solution of Me4-PACz was spin-coated, followed by 5-7 min annealing at 100 °C and IPA dynamic washing after cooling. $PbI_2$ and CsBr were then thermally evaporated, respectively, at rates of 1 Å/s and 0.1 Å/s for 1700 seconds, resulting in thin films of approximately 250 nm. The $PbI_2$/CsBr scaffold was converted to the perovskite phase via spin coating using a 0.65 M ethanol solution, of which 40% was FAI and 60% was FABr. To promote film formation, 4 mg/mL urea was included in the solution. The solution was dynamically spin-cast at 4000 rpm. The annealing step was carried at 150 °C for 15 minutes at 30% RH.

*Device fabrication*

To fabricate devices, 20 nm $C_{60}$, 5 nm BCP, and 150 nm Ag, were thermally evaporated on glass/ITO/SAM/perovskite stacks through a shadow mask, defining an active area of 0.1 $cm^2$.

*Device stability measurements*

Current–voltage (J-V) characteristics of single-junction devices were measured using a Keithley 2400 source meter under simulated 1-sun AM1.5G illumination (Sun 3000 solar simulator, Abet Technologies). Measurements were conducted in an inert $N_2$ glovebox at ambient temperature without pre-conditioning. The system was calibrated using a KG-5 filtered monocrystalline Si reference cell (Newport). Both forward and reverse scans were acquired with a 10 mV voltage step and a scan rate of 100 mV $s^{-1}$. The same source meter was used for short-term MPP tracking. The MPP tracking was performed in scanning mode with the voltage step 50 mV.

*Nano-FTIR*

Nano-FTIR measurements we conducted on device-relevant stacks (glass/ITO/SAM/perovskite) using a neaspec (attocube) s-SNOM system equipped with a broadband mid-IR laser system from Toptica. The measurements were conducted using Pt/Ir coated Arrow-type cantilevers from Nano World with 285 kHz frequency and a force constant of 42 N/m. The tapping amplitude in approach was set to 50-60 nm. The sample was illuminated with laser C (~ 1100 $cm^{-1}$ -1800 $cm^{-1}$), centered at 1333 $cm^{-1}$, and the laser intensity was in the range of 60 – 100 μW. The nano-FTIR spectra were obtained by normalizing the measured data to a silicon reference using NeaScan software. The signal was demodulated at third harmonic of the oscillation frequency of the cantilever which provides more background-free information. The nano-FTIR spectra were recorded with 10 $cm^{-1}$ spectral resolution, 25 ms per point, and 10 averages per spectrum. The nano-FTIR absorption $\alpha_n(\omega)$ was plotted as imaginary part of the complex valued near-field scattering coefficient $\sigma_n(\omega)$ according to following: $\alpha_n(\omega) = \mathrm{Im}[\sigma_n(\omega)] = s_n(\omega)\sin[\phi_n(\omega)]$. The phase tilt occurring due to the thermal drift and phase offset were corrected for using NeaPlot software. To ensure consistency of results, multiple areas of interest and multiple samples were analyzed.

*AFM*

AFM data have been collected using a Veeco Dimension Icon system with Pt/Ir coated SCM-PIT-v2 cantilevers from Bruker with 75 kHz frequency and 3 N/m force constant. The measurements were conducted on glass/ITO/SAM/perovskite samples.

*KPFM*

AFM/KPFM measurements performed using a Bruker Dimension Icon SPM. The Pt/Ir coated SCM-PIT-v2 probe with 75 kHz cantilever frequency and 3 N/m force constant was used for both topography and CPD mapping in a dual-pass mode with lift height set to 10 nm. The measurements were conducted on glass/ITO/SAM/perovskite samples. The sample was electrically grounded to the AFM sample stage with copper tape.

*Confocal PL maps*

The high-resolution PL maps were collected using a Leica SP8 microscope with a 532 nm laser system and a 93x glycerol-immersion objective lens (1.3NA). The HyD4 detector was set to the 650 nm – 795 nm range to collect emission from 1.68 eV perovskite (centered ~739 nm). The laser power was attenuated to 0.2 % (~ 80 nW). The images were collected in several frames, then drift-corrected and averaged during post-processing to enhance signal-to-noise ratio. The measurements were conducted on glass/ITO/SAM/perovskite samples.

*XRD*

XRD data were collected with Bruker D2 Phaser equipped with a 1.54184 Å X-ray source, a 1 mm slit, and a 3 mm knife. The data was collected with 0.008° step and 0.5 s exposure per point. The measurements were conducted on glass/ITO/SAM/perovskite samples.

*PL spectra*

PL spectra were collected using a confocal Raman-PL system (NT-MDT) equipped with a 473 nm laser and a 100x objective lens. The laser intensity was minimized using optical density filters to avoid sample degradation. Spectra were collected with a 300 µm slit, 1 second exposure and 10 accumulations.

*Confocal Raman spectra*

Confocal Raman spectra were collected using WITec Apyron system with 1064 nm laser and 100x Zeiss objective lens. The integration time per spectrum was set to 120 s with 5 accumulations. The sample was placed inside the sealed Linkam Scientific stage with a top quartz window that was purged with nitrogen during the measurements. The spectra were collected on glass/perovskite sample regions.

*RLM imaging*

RLM images were collected in reflection geometry using a Nikon Eclipse polarizing microscope equipped with a halogen lamp. High-resolution images were collected using a 100x Nikon objective lens (0.9 NA) yielding spatial resolution of ~ 300 nm.

*Photo- and thermal- stress experiments*

The photo-stress experiments were conducted by placing samples inside a temperature control stage (Linkam Scientific) that was gently purged with nitrogen during the experiments. The samples were illuminated through the top quartz window using a fiber-bundle-coupled tungsten-halogen lamp (Thorlabs). The intensity at the sample was set to a 3 Sun equivalent intensity to observe accelerated effects.

The thermal-stress experiments were conducted with the same temperature control stage without external illumination. The temperature was raised in 10 °C steps, each kept for about 10 minutes, until 85 °C was reached and kept for 1 hour.

The regions of interest (ROIs) were marked by carefully scratching the sample surface with tweezers. The scratches can be easily identified across different microscopes. The ROIs were pre-imaged before the stress tests and then imaged following the illumination or heat exposure. The images were post-processed in ImageJ software, including alignment and cropping to accurately trace the microscopic changes within the same ROIs.

## Data availability

Data is available from the corresponding authors upon request.

## Acknowledgements

The authors acknowledge the use of resources of the Imaging and Characterization Core Lab facilities at KAUST. This work was supported by the King Abdullah University of Science and Technology (KAUST) Research Funding Office under Award No. ORA-CRG10-2021-4681.

## Competing interests

The authors declare no competing interests.

## Author contributions

S.K. conceived the idea and carried out nano-FTIR, KPFM, PL, XRD, and RLM measurements, data analysis and interpretation, and developed CNN-based image analysis. A.R.P., V.H., S.M., L. M. and A.S.S. fabricated perovskite samples. V.H. and A.R.P. fabricated devices. V.H. carried out device stability measurements. J.P.J. carried out confocal Raman measurements and supported accelerated degradation measurements. F.L. supervised J.P.J. and contributed to scientific discussions. S.D.W. supervised S.K., A.R.P., V.H., S.M., L.M., A.S.S., provided funding, and contributed to scientific discussions. S.K. wrote the first draft of the manuscript. All authors contributed to the manuscript revision.

# Supplementary Information

## High-Throughput Imaging of Degradation-Inducing Microscopic Impurities in Perovskite Solar Cells

Sofiia Kosar[1,‡*], Anil R. Pininti[1], Vladyslav Hnapovskyi[1], José P. Jurado[1], Subhashri Mannar[1], Lorenzo Mardegan[1], Anand S. Subbiah[1], Frédéric Laquai[1,2], Stefaan De Wolf[1*]

*[1] Center for Renewable Energy and Storage Technologies (CREST), Physical Science and Engineering Division (PSE), King Abdullah University of Science and Technology (KAUST), Thuwal 23955-6900, Saudi Arabia*

*[2] Chair of Physical Chemistry and Spectroscopy of Energy Materials (SPECTRE), Department of Chemistry, Ludwig-Maximilians-University (LMU) Munich, Butenandtstraße 5-13, Munich 81377, Germany*

[‡] *Present address: Department of Chemistry and Center for Nanoscience (CeNS), Ludwig-Maximilians-University (LMU) Munich, Butenandtstraße 5-13, Munich 81377, Germany*

[*]Email: sofiia.kosar@kaust.edu.sa, stefaan.dewolf@kaust.edu.sa

### Table of Contents:

## 1. Supplementary Notes

**Supplementary Note 1. Operational principle of Nano-FTIR spectro-microscopy**

The nano-FTIR measurements are based on a s-SNOM system which combines AFM and a broadband mid-IR laser. Nano-FTIR offers a dual capability to map the surface morphology and simultaneously collect local FTIR spectra enabling surface chemical identification with a spatial resolution of several tens of nanometers[1]. Similar to regular AFM, a metal-coated tip oscillates with the mechanical resonance frequency of a cantilever (Ω) and scans the sample. Additionally, the AFM tip is illuminated with a broadband IR light using a parabolic mirror. The tip concentrates the IR light at the nanoscale spot at its apex. This gives rise to very high local fields that decay exponentially with distance. When the illuminated tip approaches the sample, the near-field interaction between tip and sample modifies the tip-scattered light. This scattered light that contains information about the local optical properties of the sample surface, is collected by the same parabolic mirror used to focus the incident light and is directed to the detector through an asymmetric Michelson interferometer where the signal is recorded as a function of the reference mirror position, yielding an interferogram. The Fourier transform of the interferogram produces amplitude $s_n(\omega)$ and phase $\varphi_n(\omega)$ signals. To plot the nano-FTIR spectra, the signal measured from the sample is normalized to the signal measured from a spectrally flat reference, e.g., silicon. The nano-FTIR absorption $\alpha_n(\omega)$ can be then plotted as the imaginary part of the complex valued near-field scattering coefficient $\sigma_n(\omega)$ according to following equation: $\alpha_n(\omega) \equiv \mathrm{Im}[\sigma_n(\omega)] = s_n(\omega)\sin[\varphi_n(\omega)]$[1]. To suppress the contributions from far-field background, the detector signal is demodulated at higher harmonics of the oscillation frequency of the cantilever. In this work, signals demodulated at 3Ω were used for analysis.

**Supplementary Note 2. Impact of film fabrication conditions on local morphology and composition**

To understand the evolution of impurity phases upon varying the fabrication conditions of our blade-coated perovskite films, and in order to test the universality of the nano-FTIR approach for impurity identification, we performed complementary nano-FTIR and XRD measurements of blade-coated films prepared with (1) varied composition and (2) bulk modifications.

*(1) Variation of composition*

We replaced A-site cation with FAMA-only, yielding Cs-free double-cation composition to understand the impact of Cs addition on residual of impurity phases in our blade-coated films. The addition of Cs is known to stabilize perovskite phase and suppress photo-inactive hexagonal phases[2]. Here, we intentionally take a reverse step inducing more hexagonal phases and probe their nanoscale distribution with nano-FTIR. Supplementary Fig. 4a shows that double-cation films form very small grains with rough morphologies and a large density of remnant impurity phases (Supplementary Fig. 4b, c) on the film surfaces. The XRD analysis shows dominance of $PbI_2$ with appreciable traces of 4H polytype phase, and only weak cubic perovskite peaks (Supplementary Fig. 4d).

*(2) Bulk modifications*

To understand the impact of bulk additives (typically used to slow down crystallization to enable growth of large grains) on morphology and local composition of triple-cation CsFAMA films, we processed films without the addition of trace amounts of PEABr, KSCN, urea and KI, while varying the annealing duration. The resulting films showed overall less-compact grains as compared to films with additives (shown in Supplementary Fig. 1). The 10 min annealing at 150 °C was found to result in incomplete small grains with rough morphologies (Supplementary Fig. 5a) and residual compositional impurities remnant on the surface (Supplementary Fig. 5b, c). XRD analysis showed strong presence of $PbI_2$ and traces of 4H perovskite polytype in addition to a weak cubic perovskite phase (Supplementary Fig. 5d). When annealing duration was extended to 20 min, perovskite films reached more compact morphologies (Supplementary Fig. 5e). The remnant compositional impurities increased in size and began to cluster (Supplementary Fig. 5f, g). The XRD showed prevalence of cubic perovskite phase and with appreciable remnant $PbI_2$ and traces of 4H polytype phase (Supplementary Fig. 5h). Further increase in annealing duration to 30 min altered the grain morphology to deviate from compact (Supplementary Fig. 6a) and resulted in increased density of compositional impurities on the film surface (Supplementary Fig. 6b, c). Structurally, the film developed strong $PbI_2$ presence with traces of 4H polytype phase (Supplementary Fig. 6d). The 40 min annealing film showed partially fragmented morphology without compact grain structure (Supplementary Fig. 6e), indicative of possible degradation. Compositionally, the films still showed presence of perovskite phase, yet with strong $PbI_2$ and 4H polytype phases (Supplementary Fig. 6f-h).

The overall smaller grains and abundant impurities for the films without bulk additives are attributed to faster crystallization in the absence of additives and high annealing temperatures required to reach cubic perovskite phase for CsFA-rich compositions.

**Supplementary Note 3. Evaluation of the U-Net model for impurity detection**

To evaluate model predictions, we introduced images of different sample types and evaluated predicted impurity masks. We first established clear correlations between chemical signature of impurities and model predictions. For this, we mapped triple-cation blade-coated perovskite films with nano-FTIR and identified compositional impurities marked with “2” and “3” in Supplementary Fig. 12a, b (initial analysis for this sample was shown in Supplementary Fig. 9). We then performed RLM imaging of the same region of interest (Supplementary Fig 12c) and introduced the RLM image to the U-Net model. Supplementary Fig. 12d shows predicted impurity mask that accurately identified impurity clusters associated with enhanced optical contrast.

We then evaluated blade-coated perovskite films prepared with different bulk modifications. Through correlative AFM/RLM measurements of triple-cation films with bulk additives, we identified areas associated with pristine grains (highlighted with grey boxes in Supplementary Fig. 13a, b) and compositional impurities, seen as hexagonally-shaped clusters (green boxes in Supplementary Fig. 13a, b). The superimposed AFM and RLM images in Supplementary Fig. 13c show clear correlation of impurity clusters in AFM with enhanced optical reflection in RLM. The impurity mask predicted by the trained U-Net model, shown in Supplementary Fig. 13d, accurately identified only compositional impurities. Similar analysis was performed for triple-cation blade-coated perovskite films that did not contain bulk additives, leading to denser impurities as marked with green boxes in Supplementary Fig. 13 e-g. The U-Net model accurately identified areas associated with impurities (Supplementary Fig. 13h).

As the model works accurately for RLM images of perovskite samples containing surface impurities (see Supplementary Fig. 14a for an example of a blade-coated film), we introduced an image of perovskite sample prepared using a hybrid method that did not result in remnant impurity phases on the film surface. Due to top-down conversion strategy, the impurity phases are expected to remain at the bottom interface for hybrid-processed perovskite films[2]. Supplementary Fig. 14b shows an RLM image of a hybrid-processed film with no surface impurities. The model accurately predicted the absence of impurity features as seen from the predicted mask, highlighting the robustness of our U-Net based image analysis.

**Supplementary Note 4. RLM analysis of impurity distribution upon bulk and surface modifications**

We employed the developed high-throughput imaging framework to diagnose the impurity distribution in our triple-cation blade-coated films upon varying the fabrication conditions as follows: upon introducing bulk (1) and surface (2) modifications.

*(1) Bulk modifications*

The bulk modifications such as removal of additives (i.e., trace amounts of KSCN, urea, PEABr and KI) from the precursor solution, as described in Supplementary Note 2, lead to a rapid reduction of nearest-neighbor distances ($NND_{I-I}$) from 1838 nm to 947 nm, accompanied by increase in total impurity coverage ($C_I$) and reduction in impurity size ($S_I$) from 1052 nm to 686 nm (Supplementary Fig. 15). While the overall $S_I$ decreases, the distribution of impurity clusters becomes denser, increasing the number of reactive sites for future degradation and highlighting the importance of controlling film crystallization to prevent clustering of remnant impurity phases.

We note that we find the $NND_{I-I}$ to be more critical when evaluating impurity distribution than the total $C_I$, as the number of impurity clusters and their close spatial distribution ultimately determine the number of reactive sites for future degradation. The denser the initial distribution of impurities, the larger the likelihood for adjacent pristine perovskite grains to degrade.

*(2) Surface modifications*

The post-treatment such as deposition of propane-1,3-diammonium iodide (PDAI) on as-prepared films - commonly used to passivate defects and improve the device $V_{OC}$ [3] - was found to result in decreased $S_I$ (448 nm as compared to 541 nm for un-passivated) but marginally impacted the $NND_{I-I}$ (833 nm as compared to 835 nm for un-passivated sample) (Supplementary Fig. 16). While the $S_I$ decreases, reducing overall $C_I$, the spatial distribution of impurities remains unchanged, maintaining the number of reactive sites for future degradation. Our results indicate that surface passivation only marginally impacts the remnant compositional impurities and is insufficient to resolve them, in contrast to reports stating that deposition of bulky ligands helps mitigate surface impurities by enabling formation of 2D phases [4]. We thus conclude that post-treatment alone is insufficient to fully resolve remnant compositional impurities.

**Supplementary Note 5. RLM analysis of impurity-induced degradation**

We monitored photo- and thermally-induced degradation of our triple-cation blade-coated perovskite films with the rapid RLM method. For this, we first imaged our samples with RLM (Supplementary Fig. 21a) and subsequently subjected them to photo-stress with 3-Sun equivalent illumination incident from the glass side - relevant geometry for single-junction perovskite solar cells. Same regions of interest were imaged following the photo-stress after 1 h, 2 h, and 3 h of illumination (Supplementary Fig. 21b-d).

To analyze changes in optical contrast, we first employed a more conventional image analysis approach, where we plotted the greyscale image intensities of same region of interest before and after photo-stress, and calculated Pearson correlation coefficients, *r*, where *r* deviating from 1 would signify lack of pixel-to-pixel intensity correlation, implying material changes. We observed that following 1 hour of photo-stress, *r* exhibited values of 0.95 and progressively dropped to 0.92 and 0.86 for 2 hours and 3 hours of photo-stress, respectively (Supplementary Fig. 21e). When analyzing intensity plots, we observed that this deviation occurred due to appearance of more pixels with lower optical reflectance (Supplementary Fig.

21f-h). Further U-Net analysis enabled us to plot accurate maps of initial impurities and photo-induced degradation shown in Supplementary Fig. i-l, demonstrating that degradation sites are anchored to the initial impurities. Following 7 hours of illumination (Supplementary Fig. 22a and b), $r$ further dropped to 0.68 (Supplementary Fig. 22c), indicative of significant alterations to the initial optical contrast, seen as propagation of degraded areas (Supplementary Fig. 22d-f).

Upon thermal stress in the absence of illumination, $r$ was also found to deviate from unity due to appearance of low-reflection contrast, showing values of 0.8 (Supplementary Fig. 23a-c). The impurity-degradation map and quantitative analysis show similar behavior as in the case of photo-stress (Supplementary Fig. 23d-f), with similar impurity-degradation nearest-neighbor distances ($NND_{I-D}$), indicating the occurrence of degraded areas adjacent to initial impurities. These findings highlight that thermal effects play a critical role in impurity-driven degradation of PSCs.

**Supplementary Note 6. Imaging techniques for stability-relevant quality-assessment of PSCs.**

Imaging approaches used to assess the quality of perovskite films generally fall into two regimes: high-resolution nanoscale mapping techniques (spatial resolutions in the range of 1 - 100s nm), and camera-based optical imaging (spatial resolutions of ~ 10s - 100s µm). High-resolution mapping methods provide detailed information about nano- and microscopic heterogeneities in morphology, chemical composition, and opto-electronic properties, whereas camera-based methods are typically employed to diagnose larger-scale defects such as scratches, cracks, and other long-range non-uniformities.

Nanoscale characterization techniques are particularly valuable for identifying microscopic degradation mechanisms and probing stability-relevant heterogeneities. Examples include (scanning) transmission electron microscopy (STEM/TEM)[5-7], scanning electron microscopy[8], scanning electron diffraction (SED)[9,10], photoemission electron microscopy (PEEM)[9,11,12], electrical modes of AFM (such as Kelvin probe force microscopy, KPFM[13-15], and conductive-AFM, c-AFM[16-18]), nano-FTIR[19], nano X-ray diffraction, nano-XRD[12], and hyperspectral photoluminescence (PL)[20,21]. These approaches provide essential chemical, structural and electronic information at nanometer length scales, and have revealed important insights about perovskite degradation. Yet their long acquisition times and specialized instrumentation generally limit measurements to small micrometer-sized regions and make them incompatible with high-throughput diagnostics or in-line process monitoring.

Conversely, camera-based imaging approaches such as photoluminescence (PL)[22-25], electroluminescence (EL)[26-28] and dark lock-in thermography (DLIT)[29,30] enable rapid imaging of large device areas and therefore show strong potential for high-throughput characterization. However, these methods primarily probe optoelectronic responses and typically lack the spatial resolution and chemical specificity needed to directly identify microscopic stability-relevant features, including compositional impurities and generally remain uncalibrated for early stability assessment. Moreover, EL and DLIT imaging are commonly performed on complete device stacks, limiting their applicability for pre-screening film quality prior to device fabrication.

High-throughput RLM combined with CNN-based image analysis introduced in this work provides a balance between spatial resolution, image acquisition speed and stability relevance. The spatial resolution (~100s nm) is sufficient to directly detect microscopic compositional impurities that act as degradation initiation sites in perovskite films. Importantly, this optical contrast is chemically validated through correlative RLM and nano-FTIR measurements, enabling reliable identification of compositional impurities. Image acquisition and automated analysis require only a few seconds, allowing rapid quantification of impurity density in as-prepared films. When applied prior to device completion, this approach enables non-invasive pre-screening of film quality and evaluation of susceptibility to impurity-driven degradation. Because both imaging and analysis operate on second timescales and rely on simple optical hardware, the method is compatible with high-throughput workflows and can be integrated into in-line characterization during device fabrication without introducing delays. As such, the rapid RLM framework provides a practical intermediate characterization step that bridges nanoscale stability-relevant microscopic insights and manufacturing-compatible diagnostics for fabrication of stable and scalable PSCs.

A comparison of commonly used imaging and mapping techniques with respect to spatial resolution, acquisition speed, sample compatibility, potential for in-line integration and relevance to stability assessment is summarized in Supplementary Table 1.

**Supplementary Table 1. Comparison of imaging and mapping methods relevant for assessing perovskite film quality.**

| Method | Spatial resolution | Acquisition time per image | Damage to sample | Sample compatibility | In-line potential | Stability diagnostic capability | Ref. |
|---|---|---|---|---|---|---|---|
| TEM/STEM | < 1 nm | minutes to hours | no/moderate | thin lamella/film on TEM grid | low | yes | 5-7 |
| SEM | few nm | minutes | yes | film/device-relevant stack | low | yes | 8 |
| SED | 10s nm | minutes to hours | no/moderate | film/device-relevant stack | low | yes | 9,10 |
| PEEM | 10s nm | minutes | no | device-relevant stack | low | yes | 9,11,12 |
| KPFM/c-AFM | 10s nm | seconds to hours | no | film/device-relevant stack | low/ moderate | yes | 13-15 |
| Nano-FTIR | 10s nm | minutes to hours | no | device-relevant stack | low | yes | 19 |
| Nano-XRD | 10s-100s nm | minutes to hours | no | film on grid | low | yes | 12 |
| Hyperspectral PL | 100s nm | minutes to hours | no | film/device-relevant stack/device | moderate | yes | 20,21 |
| High-resolution RLM + CNN | 100s nm | seconds | no | film/device-relevant stack/device | high | yes | this work |
| Camera-based PL | 10s-100s μm | seconds | no | film/device-relevant stack/device | high | not calibrated | 22-25 |
| Camera-based EL | 10s-100s μm | seconds | no | device | high | not calibrated | 26-28 |
| Camera-based DLIT | 10s-100s μm | seconds | no | device | high | not calibrated | 29,30 |

## 2. Supplementary Figures

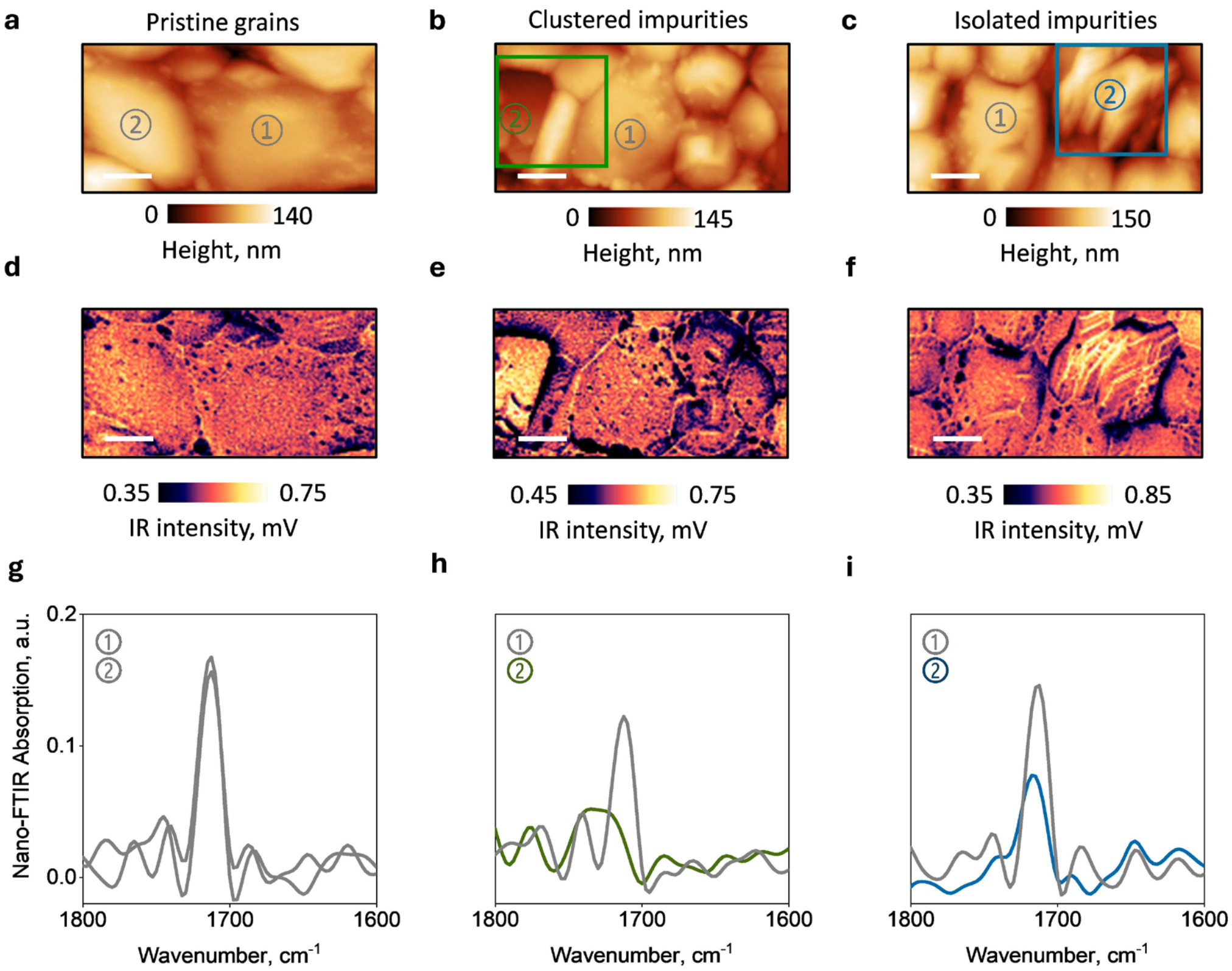

**Supplementary Fig. 1. Identification of nanoscale compositional impurities.** AFM maps of regions with (a) pristine grains, (b) hexagonally-shaped clustered impurities (marked with a green box), and (c) with an isolated impurity (marked with a blue box). (d-f) White-light IR maps corresponding to the regions in (a-c), showing strong and uniform IR absorption by the pristine grains, and weak IR absorption by the clustered and isolated impurities. (g-i) Nano-FTIR spectra collected from points of interest as indicated in AFM images in (a-c). Pristine perovskite grains are marked with grey, hexagonal clustered impurities with green, and an isolated impurity with blue. Scale bars are 500 nm.

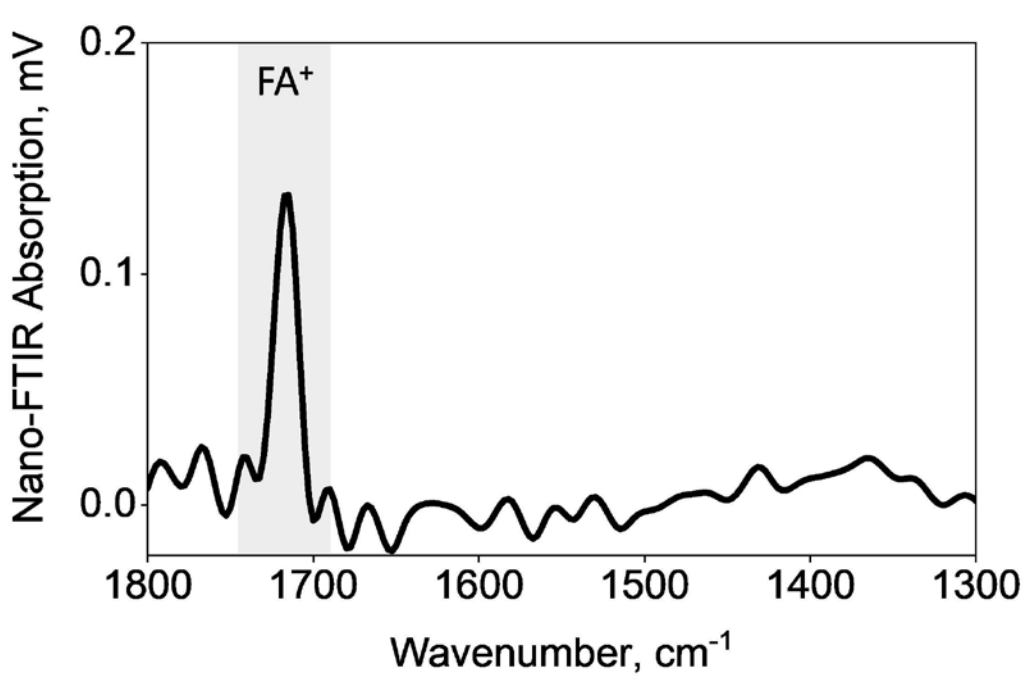

**Supplementary Fig. 2. Nano-FTIR spectrum of pristine perovskite.** The nano-FTIR spectrum of a pristine grain in the spectral range from 1300 $cm^{-1}$ – 1800 $cm^{-1}$, showing no additional resonances except for 1714 $cm^{-1}$, attributed to C=N vibration of $FA^+$ cation. No obvious peaks were detected that can be attributed to $MA^+$ cation or residual solvents.

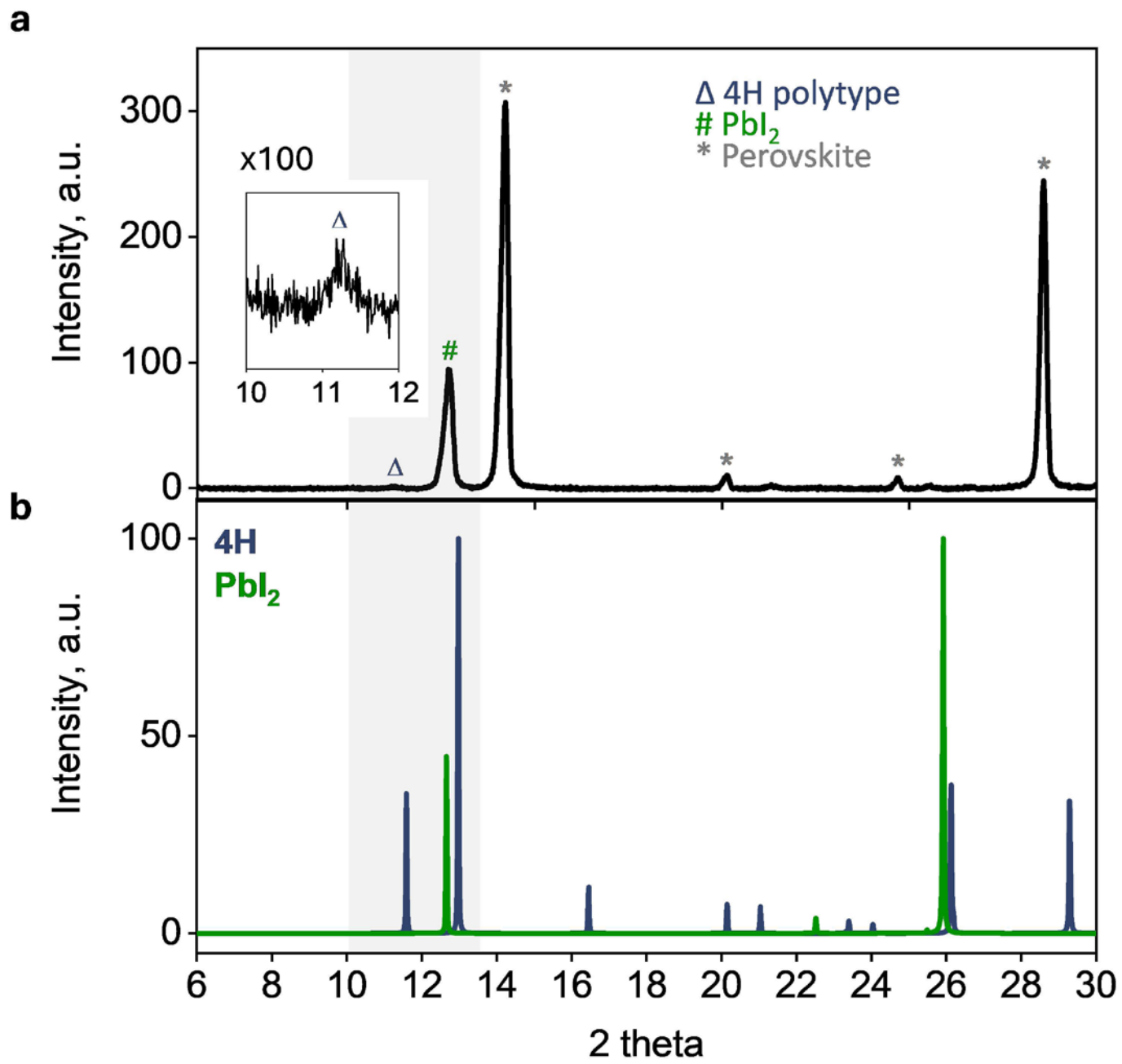

**Supplementary Fig. 3. XRD of a blade-coated perovskite film.** (a) XRD of perovskite film showing the presence of a dominant cubic perovskite phase (marked with "*"), and inclusions of remnant $PbI_2$ (marked with "#") and 4H hexagonal perovskite polytype phase (marked with "Δ"). Magnified region highlighting 4H polytype phase is shown in the inset. (b) Modeled XRD of $PbI_2$ and 4H phase. The powder XRD spectra were modeled in Vesta software using crystal structure (.cif) files from literature references[31,32].

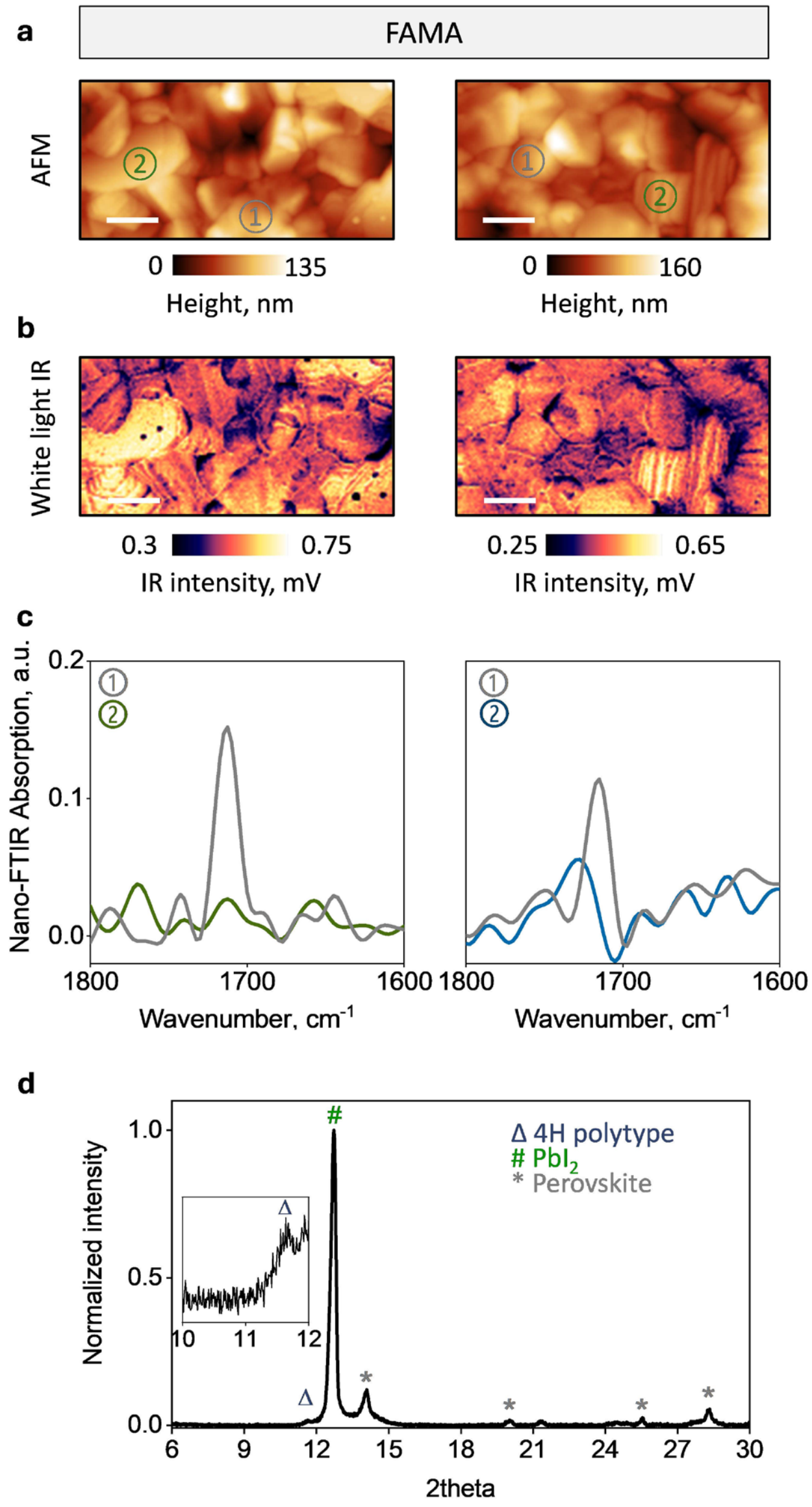

**Supplementary Fig. 4. Impact of A-site cation variation on morphology and nanoscale chemical composition of perovskite films.** (a) AFM maps of two representative regions of a double-cation FAMA film. (b) White-light IR maps corresponding to the region in (a), showing weak IR absorption by impurity features. (c) Nano-FTIR spectra collected from points of interest as indicated in AFM image in (a). Pristine perovskite grains are marked with grey and impurity features with green and blue. (d) XRD of FAMA perovskite film with magnified regions highlighting presence of remnant 4H phase. The cubic perovskite phase is marked with "*", $PbI_2$ with "#", and 4H with "Δ". Scale bars are 500 nm.

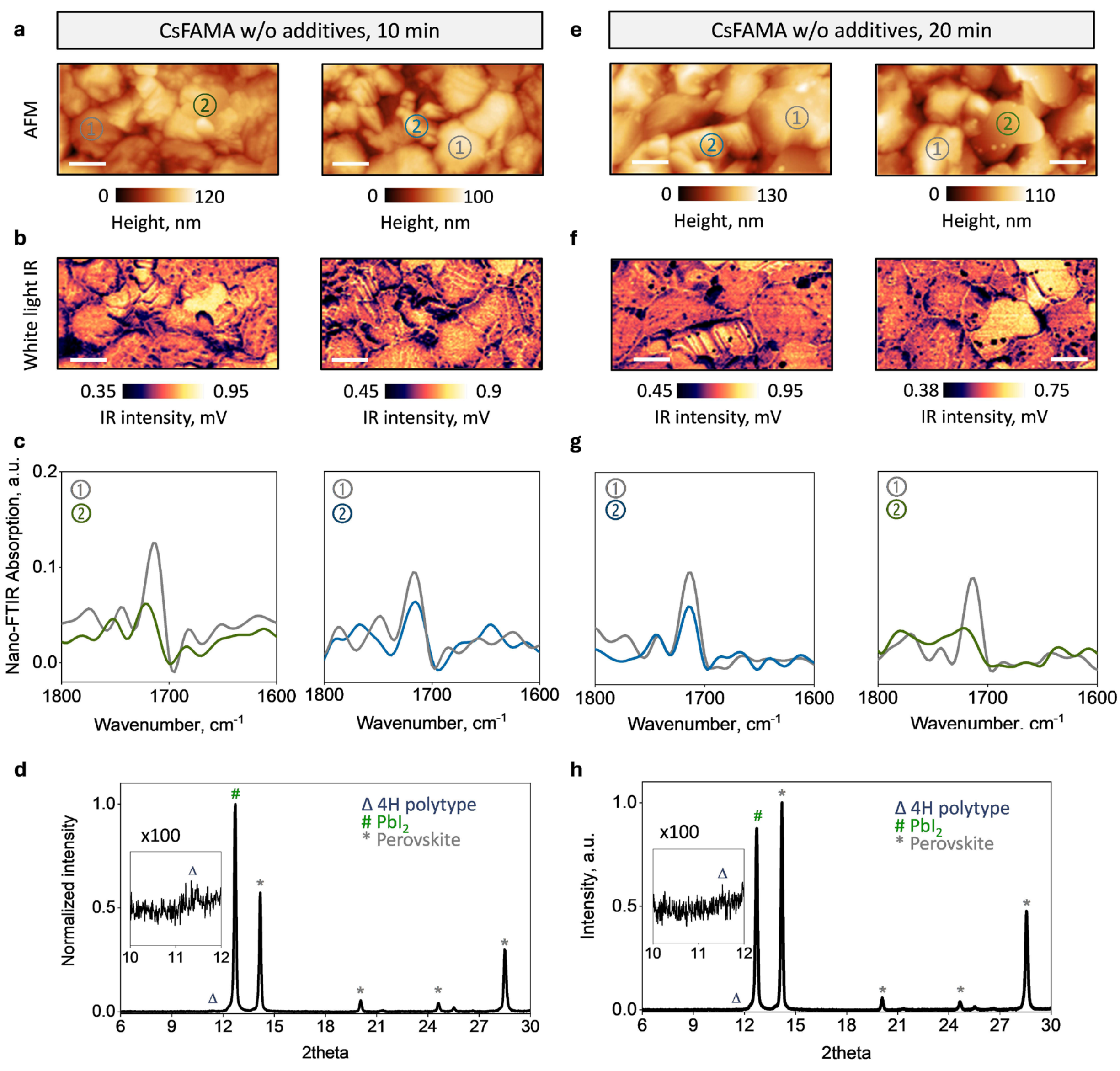

**Supplementary Fig. 5. Impact of additive modification and annealing duration (10 and 20 min).** (a, e) AFM maps of two representative regions for each film (10 min and 20 min annealing). (b, f) White-light IR maps corresponding to the regions in (a, e), showing weak IR absorption by impurity features. (c, g) Nano-FTIR spectra collected from points of interest as indicated in AFM images in (a, e). Pristine perovskite grains are marked with grey and impurity features with green and blue. (d, h) XRD of 10 min and 20 min perovskite films with magnified regions highlighting presence of remnant 4H phase. The cubic perovskite phase is marked with "*", $PbI_2$ with "#", and 4H with "Δ". Scale bars are 500 nm.

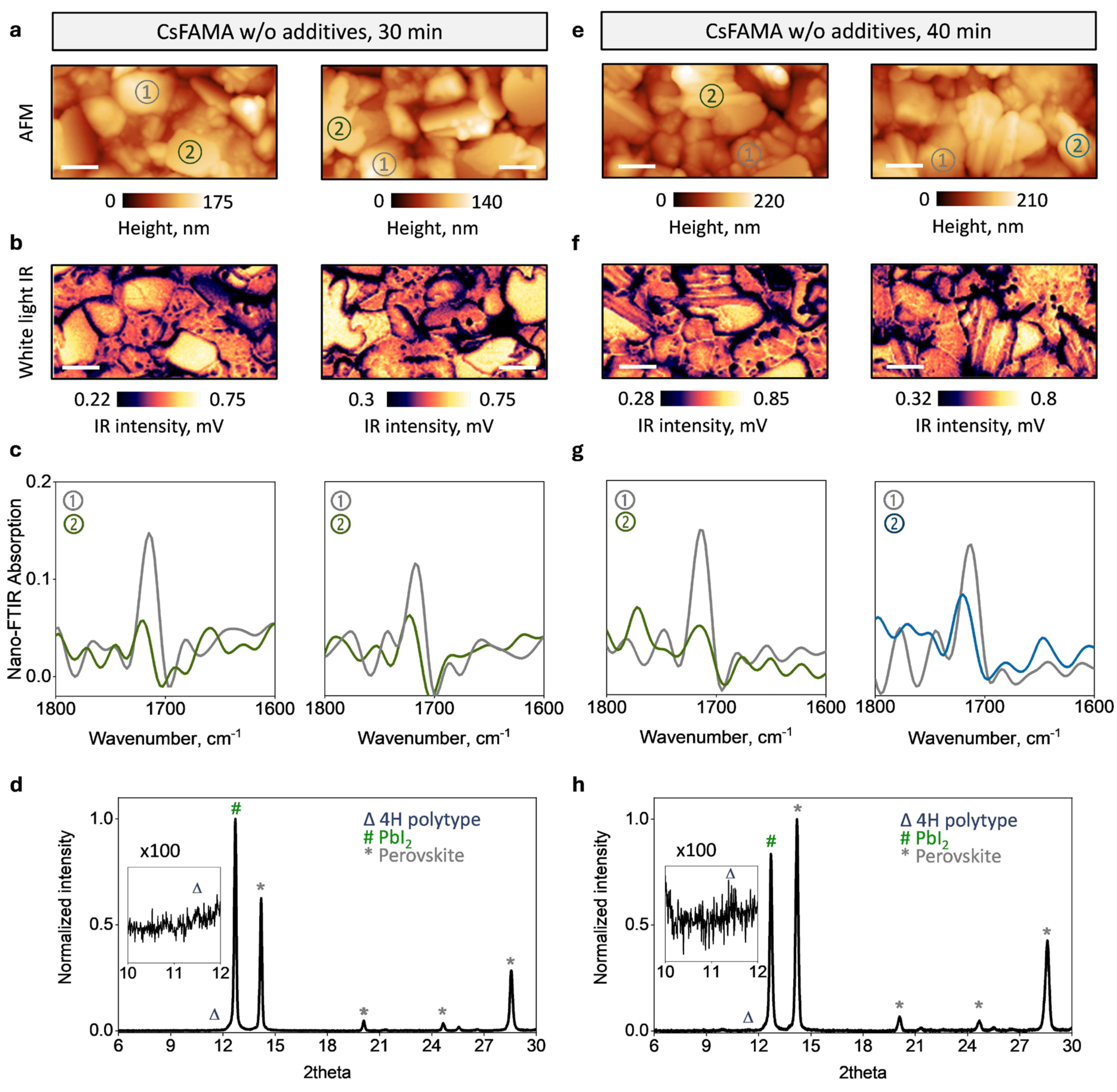

**Supplementary Fig. 6. Impact of additive modification and annealing duration (30 and 40 min).** (a, e) AFM maps of two representative regions for each film (30 min and 40 min annealing). (b, f) White-light IR maps corresponding to the regions in (a, e), showing weak IR absorption by impurity features. (c, g) Nano-FTIR spectra collected from points of interest as indicated in AFM images in (a, e). Pristine perovskite grains are marked with grey and impurity features with green and blue. (d, h) XRD of 30 min and 40 min perovskite films with magnified regions highlighting presence of remnant 4H phase. The cubic perovskite phase is marked with "*", $PbI_2$ with "#", and 4H with "Δ". Scale bars are 500 nm.

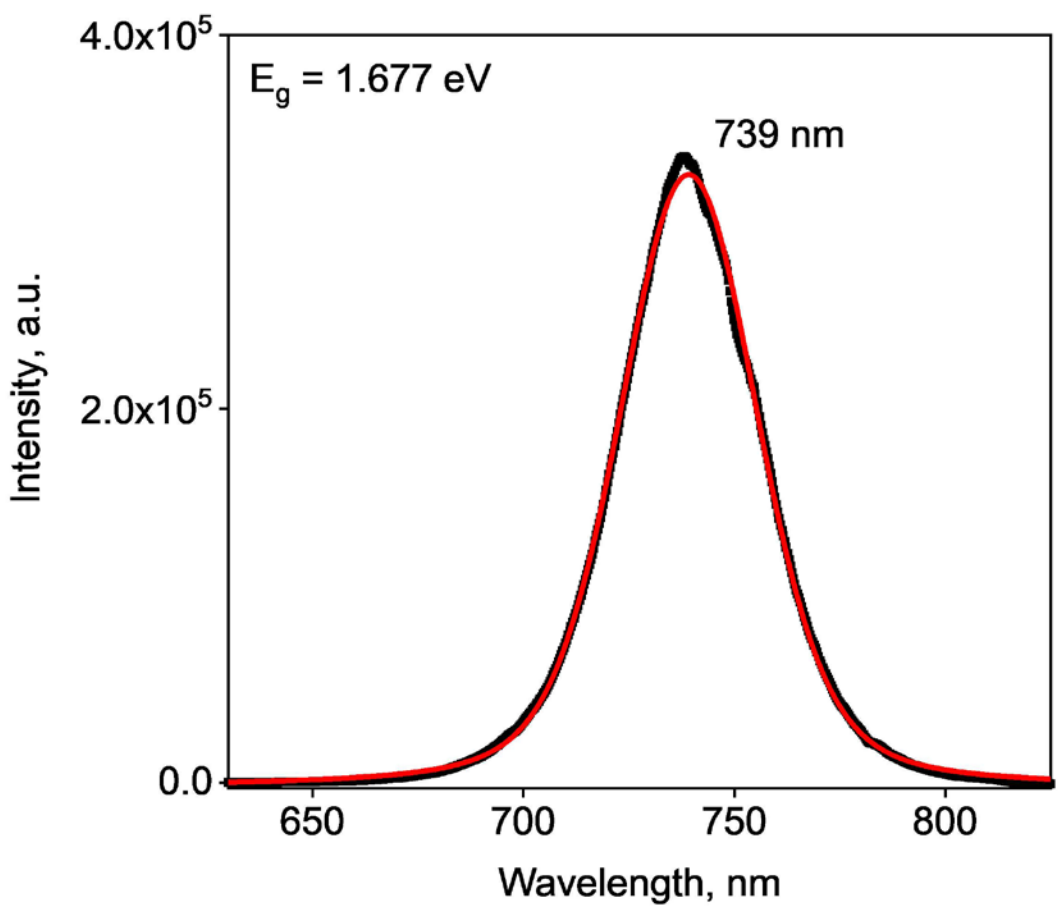

**Supplementary Fig. 7. Confocal PL spectrum of a blade-coated perovskite film.** PL spectrum (scatter) with a Voigt fit (red line), showing the PL peak at 739 nm (~1.68 eV).

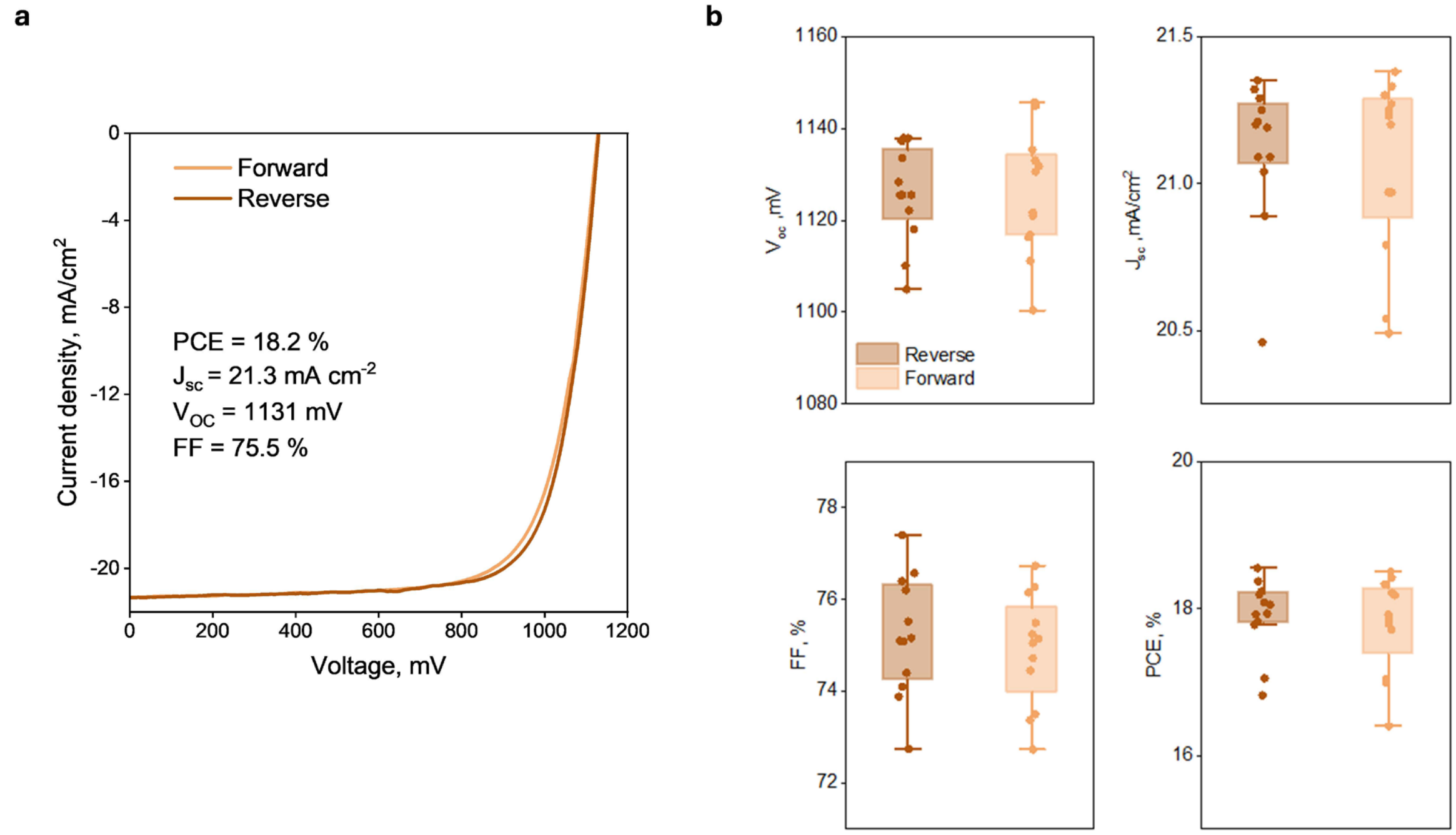

**Supplementary Fig. 8. Device performance measurements.** (a) J-V scan of a typical single-junction device based on a blade-coated perovskite film. (b) Statistics of device performance for several representative devices with forward and reverse scan data.

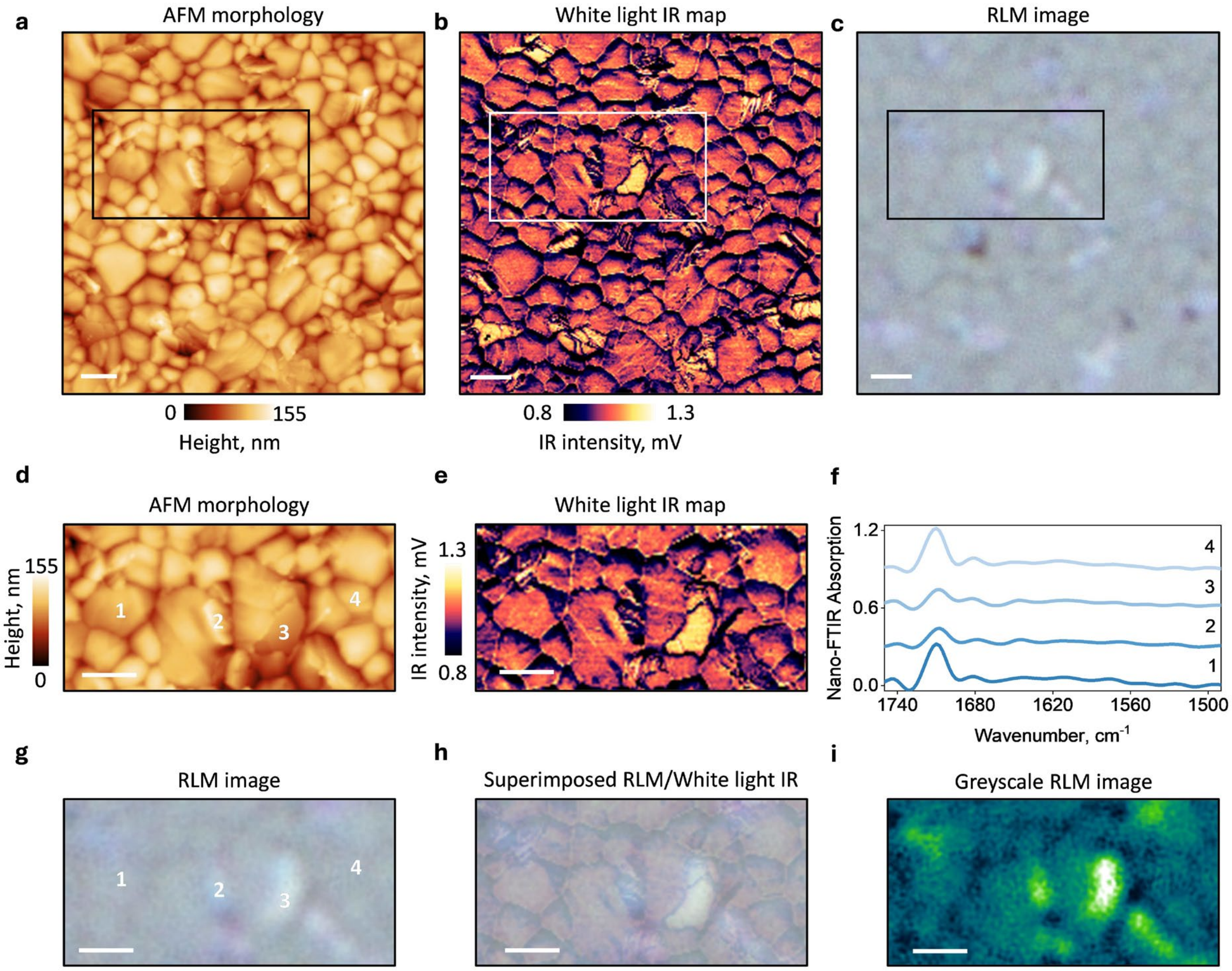

**Supplementary Fig. 9. RLM identification of compositional impurities.** (a) AFM-, (b) white-light IR-, and (c) RLM images of the same large region of interest of a blade-coated perovskite film. Magnified (d) AFM- and (e) white-light IR images of a smaller region marked with black rectangles in (a, b). (f) Nano-FTIR spectra collected from features marked "1"-"4" in the AFM image in (d), representing pristine perovskite grains (marked with "1" and "4") and impurity features without organic cations (marked with "2" and "3"). (g) RLM image of the magnified region of interest as marked in (a-c) with features of interest indicated. (h) Superimposed RLM image and white-light IR map of the same magnified region as in (d, e, g), showing that compositional impurities with weak IR absorption correlate well with enhanced optical reflectance. (i) RLM image from the same region shown in greyscale. Scale bars are 1 µm.

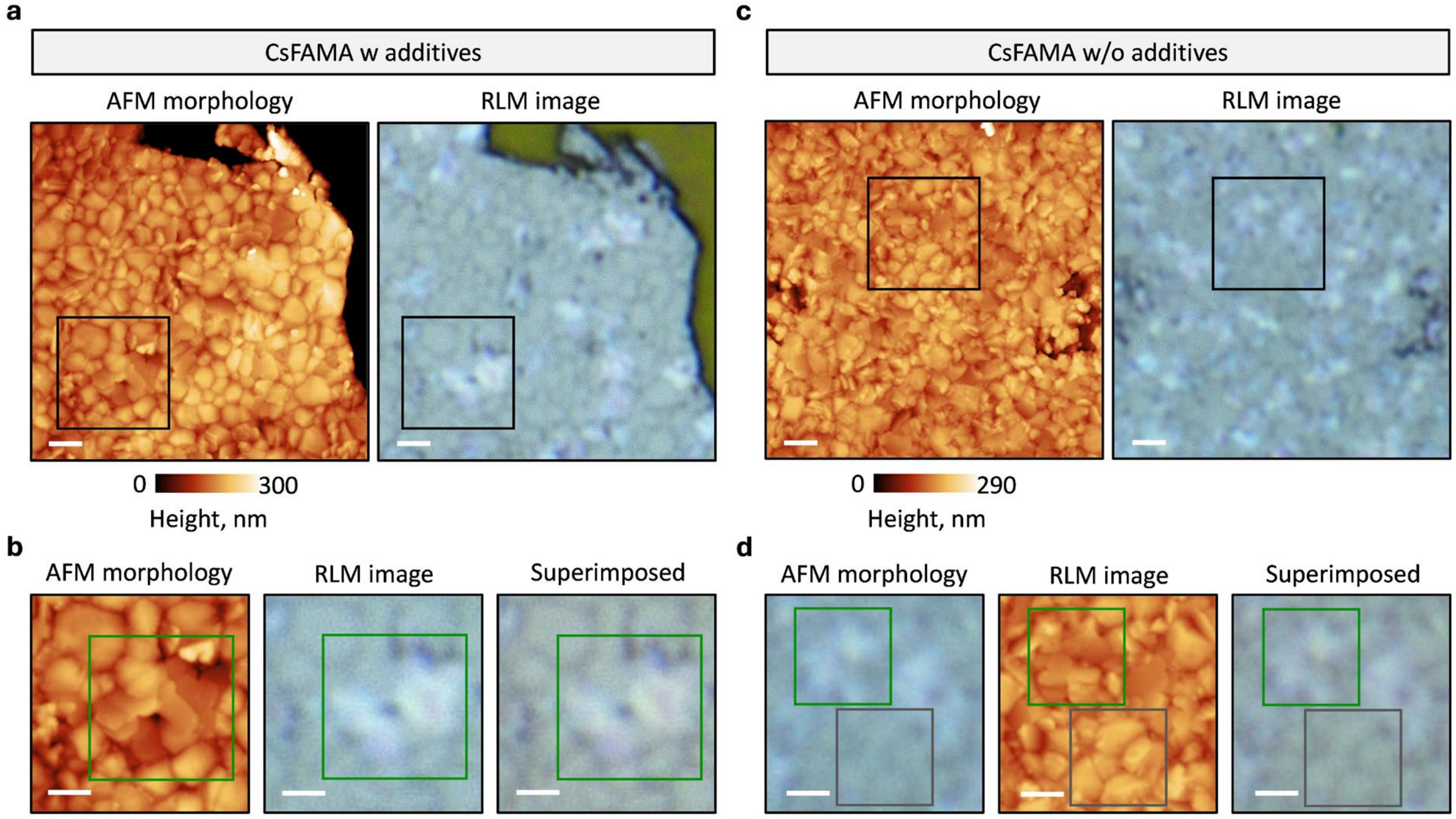

**Supplementary Fig. 10. AFM-RLM correlations for perovskite films with varied additive content.** (a) AFM and RLM images of a large region of interest of a triple cation perovskite film with additives. (b) Magnified AFM-, RLM- and their superimposed images of a smaller region of interest as marked with black boxes in (a). The impurity clusters are highlighted with green boxes. (c) AFM and RLM images of a large region of interest of a triple cation film without additives showing denser impurity distribution. (d) Magnified AFM-, RLM- and their superimposed images of the smaller region of interest as marked with black boxes in (c). The impurity clusters are highlighted with green boxes, and pristine regions with grey boxes. Only impurities are associated with enhanced optical contrast. Scale bars are 1 μm.

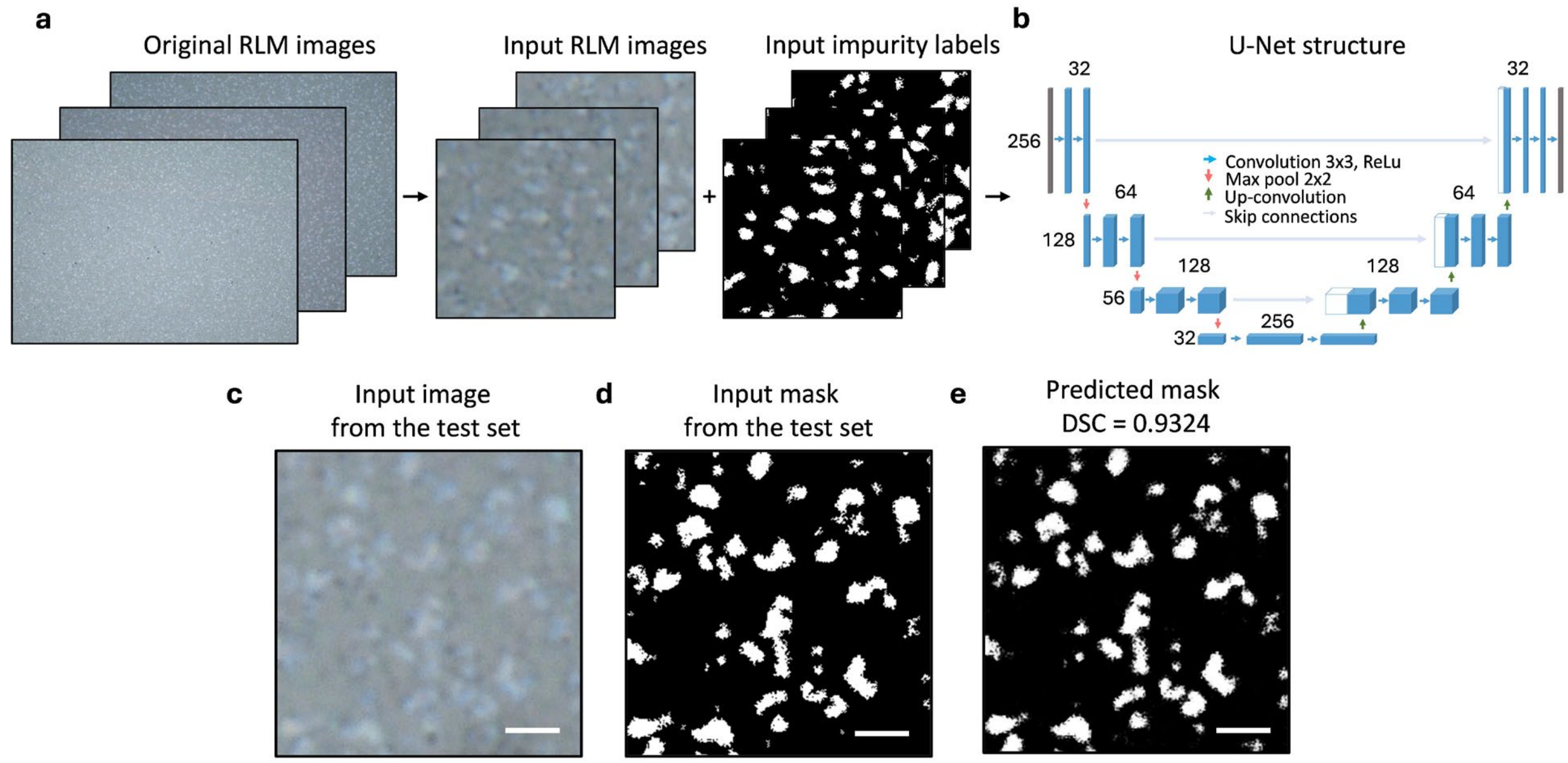

**Supplementary Fig. 11. Training and testing the U-Net model for impurity detection.** (a) Example of the training & testing set preparation: original RLM images from the microscope are cropped to smaller sizes to reduce computational load. The impurity labels are prepared by thresholding grey-scale images until only impurities with enhanced contrast remain highlighted. The images are divided 80 % for the training set, 20 % for the testing set and are used to train the model consisting of 3 encoder-decoder layers. (b) Schematic of U-Net model. (c) An input image from the test set, (d) input mask from the test set, and (e) impurity mask predicted by the model with DSC coefficient. The predicted mask accurately follows the input mask and original impurity distribution. Scale bars are 2 µm.

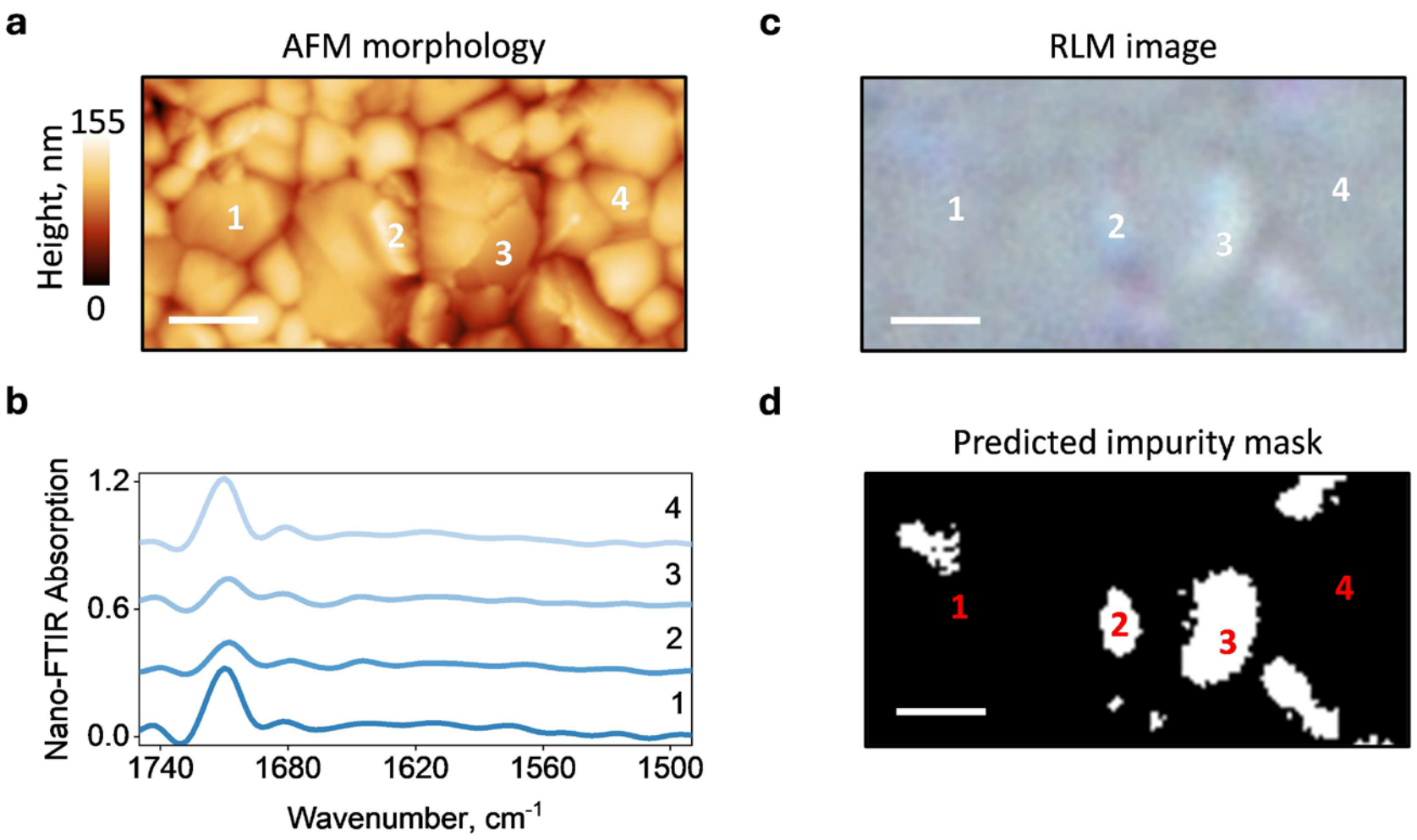

**Supplementary Fig. 12. Nano-FTIR-RLM correlations with predicted impurity masks.** (a) AFM image of perovskite film (shown initially in Supplementary Fig. 9). (b) Nano-FTIR spectra from features marked in (a-c), representing pristine grains (marked with "1" and "4"), and compositional impurities (marked with "2" and "3"). (c) RLM image of the same region as in (a). (d) Predicted impurity mask. Correlation of nano-FTIR spectra with predicted impurity mask validates the accuracy of the model predictions. Scale bars are 1 µm.

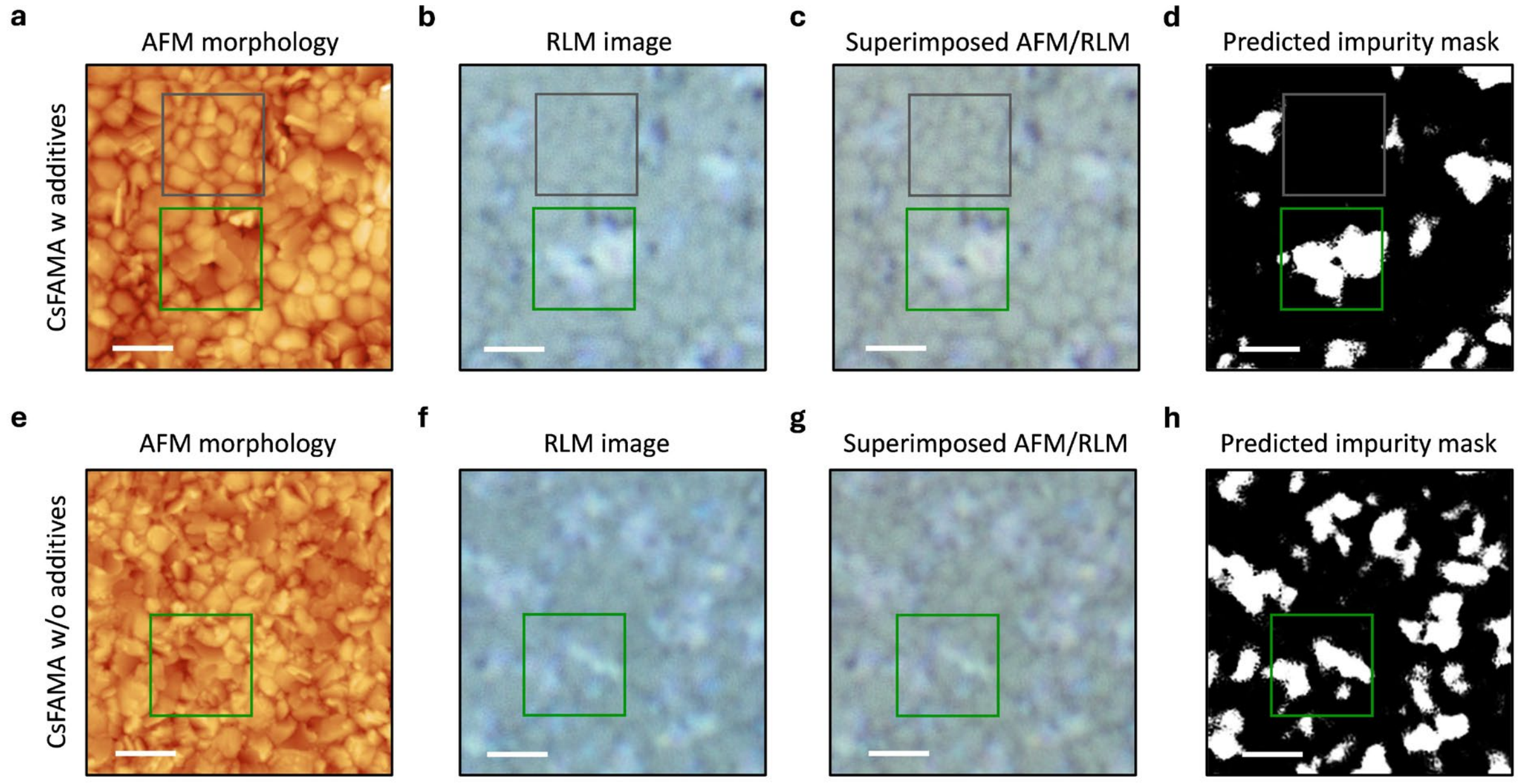

**Supplementary Fig. 13. RLM-AFM correlations for perovskite films with predicted impurity masks.** (a) AFM-, (b) RLM, and (c) superimposed AFM/RLM images of a triple-cation blade-coated perovskite film with bulk additives, showing large impurity clusters on the film surface. (d) Predicted impurity mask for the same region as in (a-c). Areas with impurity clusters are highlighted with green boxes, while pristine areas with compact grains that are not associated with enhanced optical signal are highlighted with grey boxes. (e) AFM-, (f) RLM, and (g) superimposed AFM/RLM images of a triple-cation blade-coated perovskite film without bulk additives, showing a denser distribution of impurity clusters on the film surface. (h) Predicted impurity mask for the same region as in (e-g), with impurity clusters highlighted with green boxes. Scale bars are 2 µm.

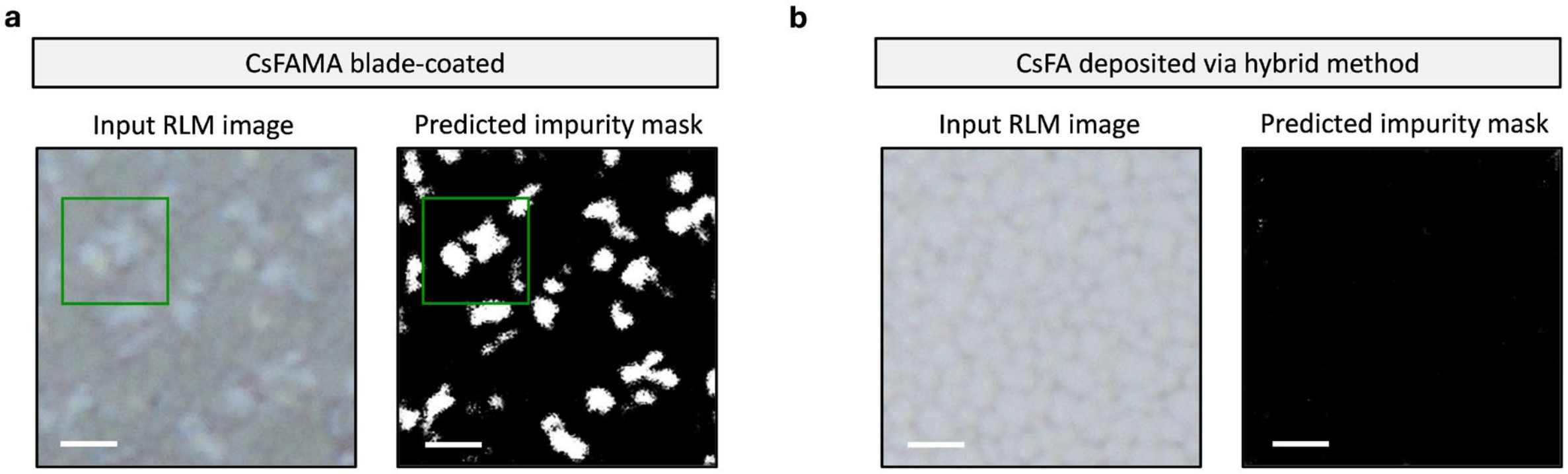

**Supplementary Fig. 14. RLM images of perovskite films with predicted impurity masks.** (a) RLM image of a triple-cation blade-coated perovskite film with multiple impurity clusters on the surface, previously unseen by the model with predicted impurity mask. The impurity clusters associated with enhanced optical contrast are marked with green boxes. (b) RLM image of a double-cation (CsFA) hybrid-processed perovskite film without any impurity clusters on the surface, previously unseen by the model with predicted impurity mask. The model accurately identified the absence of impurity features. Scale bars are 2 µm.

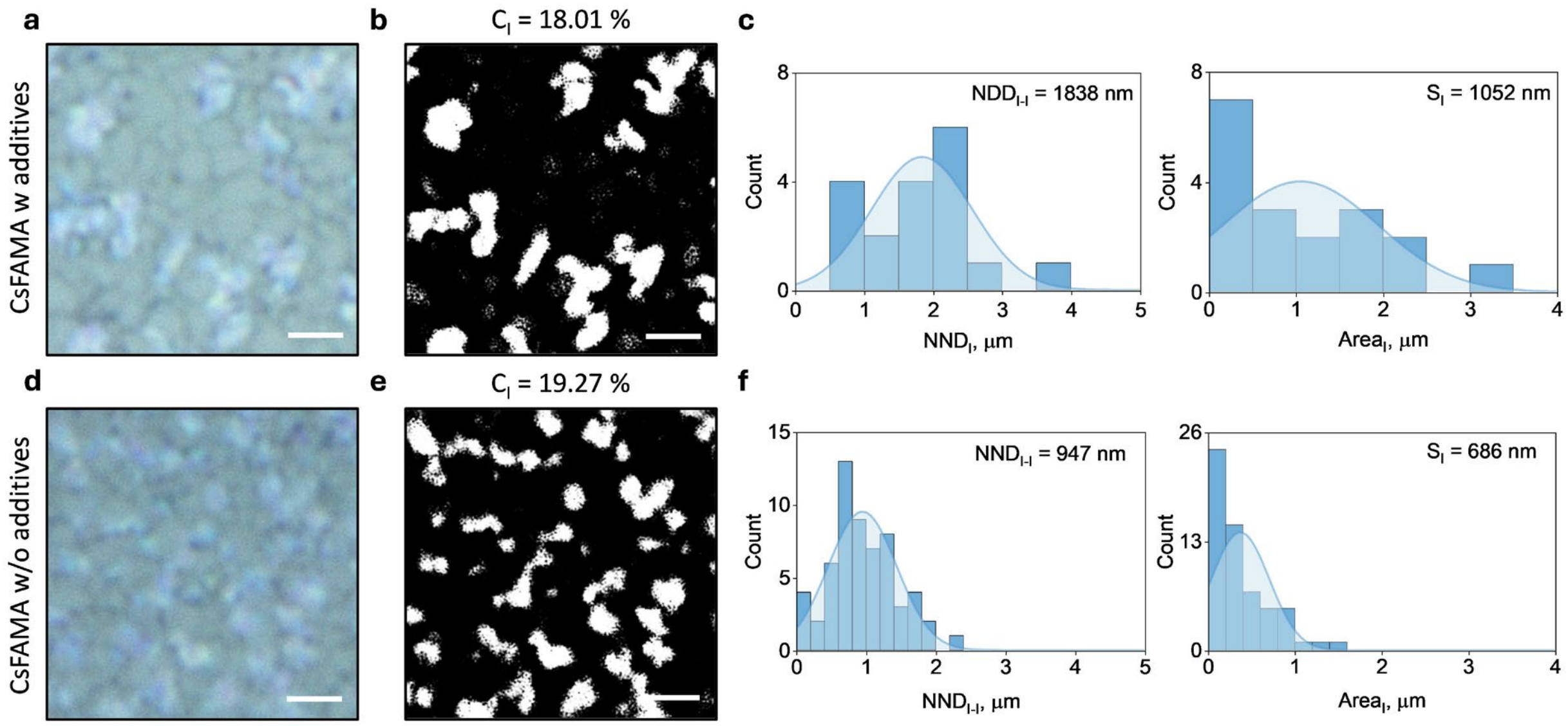

**Supplementary Fig. 15. RLM analysis of perovskite films with bulk modifications.** (a) RLM image, and (b) predicted impurity mask of a triple-cation blade-coated perovskite film with additives. (c) Nearest neighbor distance ($NND_{I-I}$) and size ($S_I$) distributions for the image shown in (a). (d) RLM image, and (e) predicted impurity mask of a triple-cation blade-coated perovskite film without additives. (f) $NND_{I-I}$ and $S_I$ distributions for the image shown in (d), highlighting denser impurity clustering in the absence of bulk additives with overall smaller impurity sizes. Impurity clusters were assumed to be circular for $S_I$ calculation. Scale bars are 2 µm.

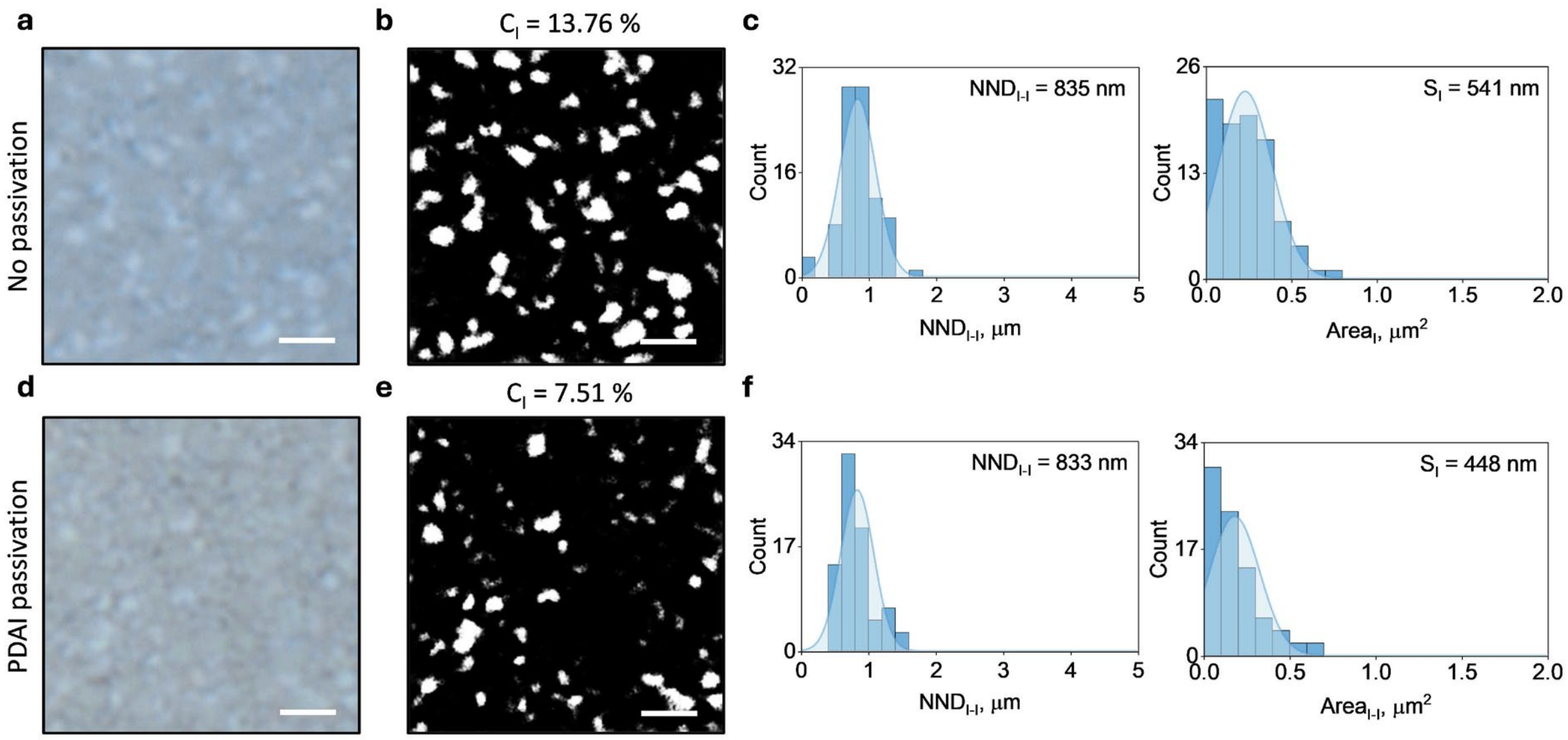

**Supplementary Fig. 16. RLM analysis of perovskite films with surface passivation.** (a) RLM image, and (b) predicted impurity mask of a triple-cation blade-coated perovskite film. (c) $NND_{I-I}$ and $S_I$ distributions for the image shown in (a). (d) RLM image, and (e) predicted impurity mask of a triple-cation blade-coated perovskite film with surface PDAI passivation. (f) $NND_{I-I}$ and $S_I$ distributions for the image shown in (d), highlighting that top passivation leads to decreased impurity sizes and marginally increased $NND_{I-I}$, yet remains insufficient to fully resolve remnant compositional impurities on the surfaces of perovskite films. Scale bars are 2 µm.

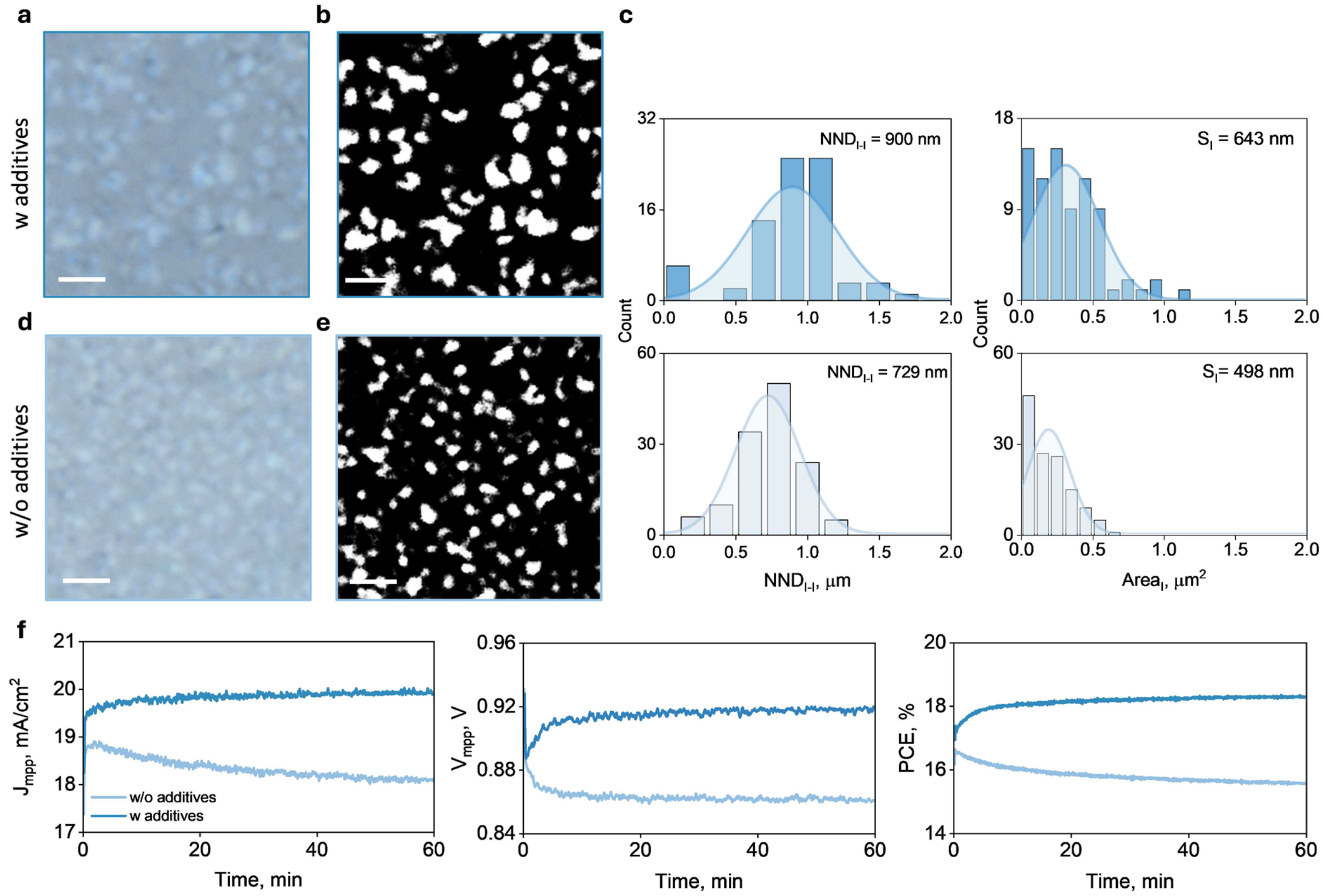

**Supplementary Fig. 17. Device stability as a function of impurity distribution.** (a) RLM image, and (b) predicted impurity mask of a triple-cation blade-coated perovskite film with bulk additives. (c) $NND_{I-I}$ and $S_I$ distributions for the images shown in (a, d). (d) RLM image, and (e) predicted impurity mask of a triple-cation blade-coated film without bulk additives resulting in denser impurity distribution. (f) Maximum power point (MPP) tracking of single-junction devices based on films with and without additives. While the device based on a film with bulk additives remains stable during the first hour of operation, the device based on the film without bulk additives shows a rapid and continuous degradation of $J_{SC}$ and a rapid and eventually slower degradation of $V_{OC}$. This correlates well with the denser impurity distribution in the case of the film without bulk additives. Scale bars are 2 µm.

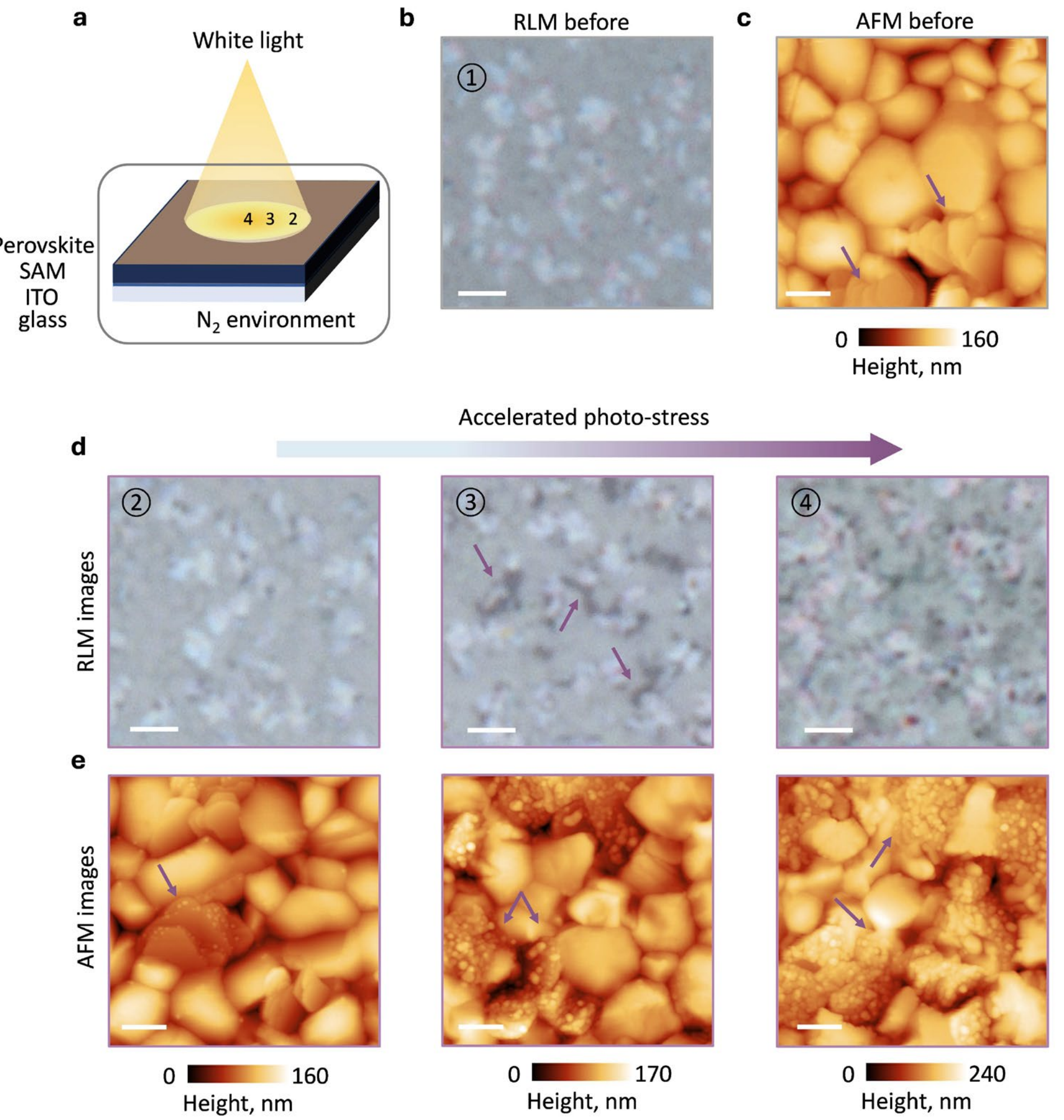

**Supplementary Fig. 18. Microscopic monitoring of degradation with RLM and AFM.** (a) Schematic of the photo-stress experiment: device-relevant stacks glass/ITO/SAM/perovskite inside the sealed stage gently purged with nitrogen are illuminated from the film side with white light. (b) RLM image before exposure showing film surface with multiple compositional impurities. (c) AFM image before exposure with impurity clusters indicated with purple arrows. (d) RLM images collected along the diverging incident beam, representing progression of photo-degradation. The degradation is seen as local darkening of optical contrast that is initially localized at impurity-perovskite junctions (indicated with purple arrows). (e) AFM images of from the same regions as in (d), showing a more microscopic picture of degradation: upon illumination, the impurity clusters begin fragmenting by developing nanoscale speckles; decomposition then propagates to impurity-perovskite junctions and eventually to the perovskite grains (indicated with purple arrows). Scale bars are 2 µm for RLM images and 500 nm for AFM images.

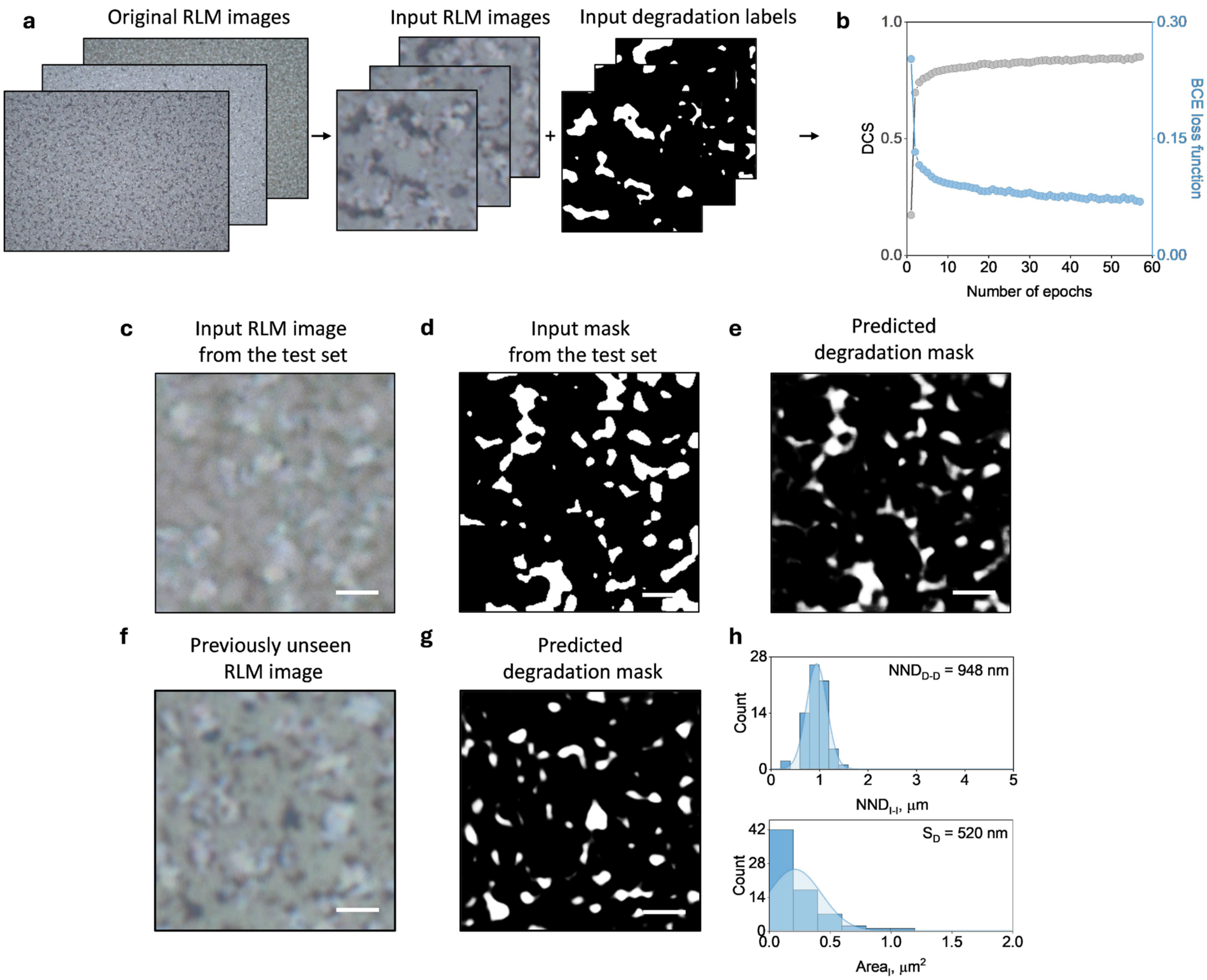

**Supplementary Fig. 19. Training and testing the U-Net model for degradation detection.** (a) Example of the training & testing set preparation: original RLM images from the microscope are cropped to smaller sizes. The degradation labels are prepared by inverse-thresholding grey-scale images until only degraded areas with reduced contrast remain highlighted. The images are divided 80 % for the training set, 20 % for the testing set and are used to train the model. (b) BCE loss function and DSC indicating effectiveness of the training. (c) An input image from the test set, (d) input mask from the test set, and (e) impurity mask predicted by the model. The predicted mask accurately follows the input mask and original degradation distribution. (f) RLM image of perovskite sample after photo-stress, previously unseen by the model. (g) Predicted degradation mask. (h) Nearest neighbor distances between degraded areas ($NND_{D-D}$) and sizes of degraded areas ($S_D$) calculated for (f). Degraded areas were assumed to be circular for $S_D$ calculation. Scale bars are 2 μm.

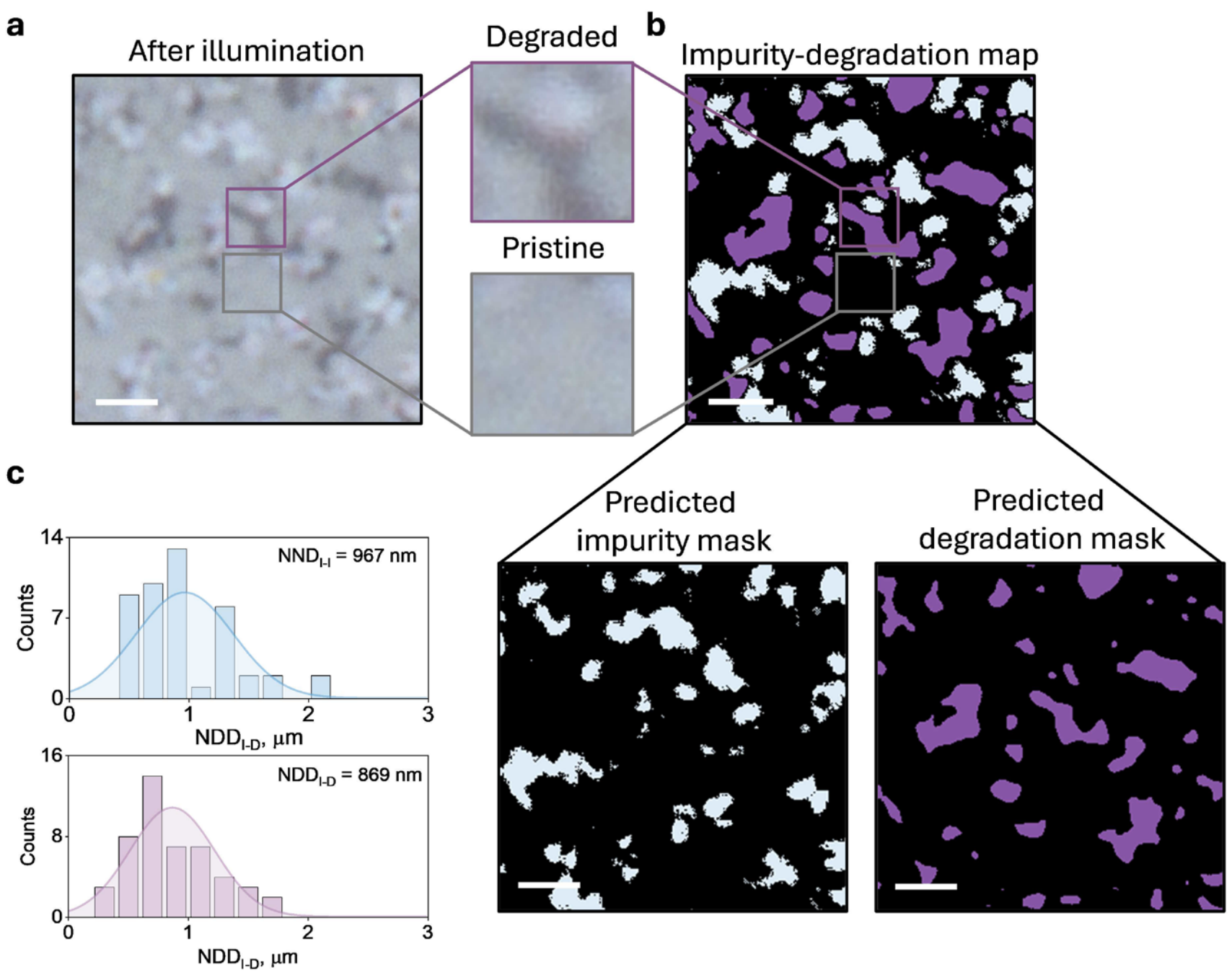

**Supplementary Fig. 20. Optical monitoring of degradation following film-side illumination.** (a) RLM image of perovskite film after photo-stress with 3 sun equivalent intensity with magnified areas showing degradation seen as reduced optical contrast at the impurity-perovskite junctions (purple box) and remaining pristine film in the area not associated with impurities (grey box). (b) Impurity-degradation map constructed by superimposing impurity and degradation masks predicted by respective U-Net models. (c) Nearest neighbor distances among impurities ($NND_{I-I}$) and impurity-to-degradation nearest neighbor distances ($NND_{I-D}$) for (a), indicating that degradation occurs in the areas adjacent to the initial impurities. Scale bars are 2 µm.

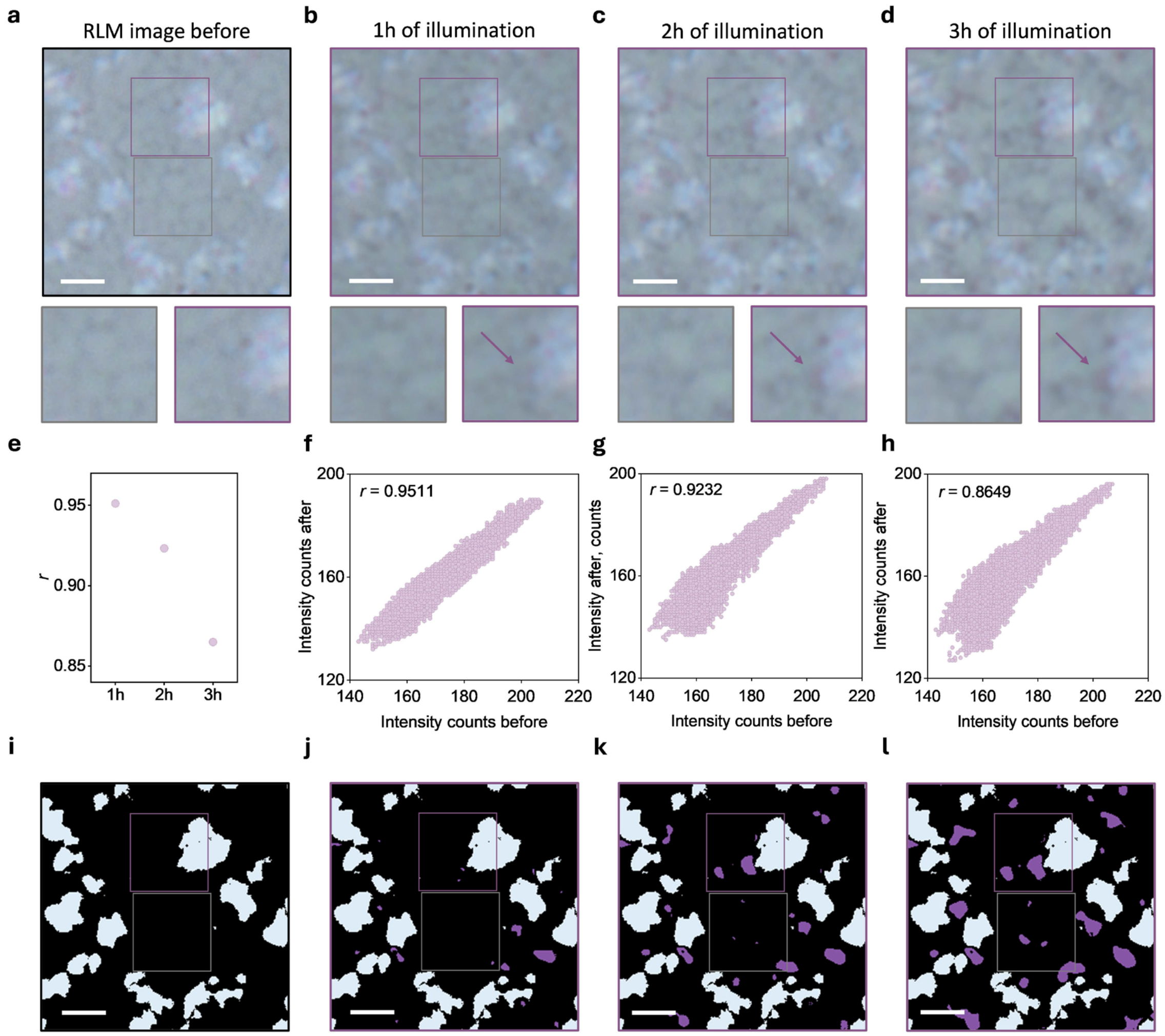

**Supplementary Fig. 21. Optical monitoring of degradation following glass-side illumination.** (a) RLM image of perovskite film before photo-stress with two regions highlighted: pristine without impurities (grey box) and impurity-perovskite junction (purple box). Magnified images of the selected regions are shown below. (b) RLM image of the same area following 1 hour of photo-stress with 3 sun equivalent intensity. Magnified images show beginning of local darkening of the optical contrast at the impurity-perovskite junction (shown with purple arrow) while pristine region remains intact. (c) RLM image of the same area following 2 hours of photo-stress. The optical darkening at the impurity-perovskite junction progresses as shown in the magnified image while pristine region remains intact. (d) RLM image of the same area following 3 hours of photo-stress. The impurity-perovskite junction shows clear darkening of optical contrast. Some grain boundaries in the pristine region begin showing darkening. (e) Pearson correlation coefficient (*r*) for the regions in (b-d). (h) RLM image intensity correlations before-after degradation for (b-d) with calculated *r*. (i) Predicted impurity mask for (a). (j-l) Impurity-degradation maps for (b-d) showing initial distribution of impurities and progressive appearance of degradation localized adjacent to the initial impurity features. Scale bars are 2 μm.

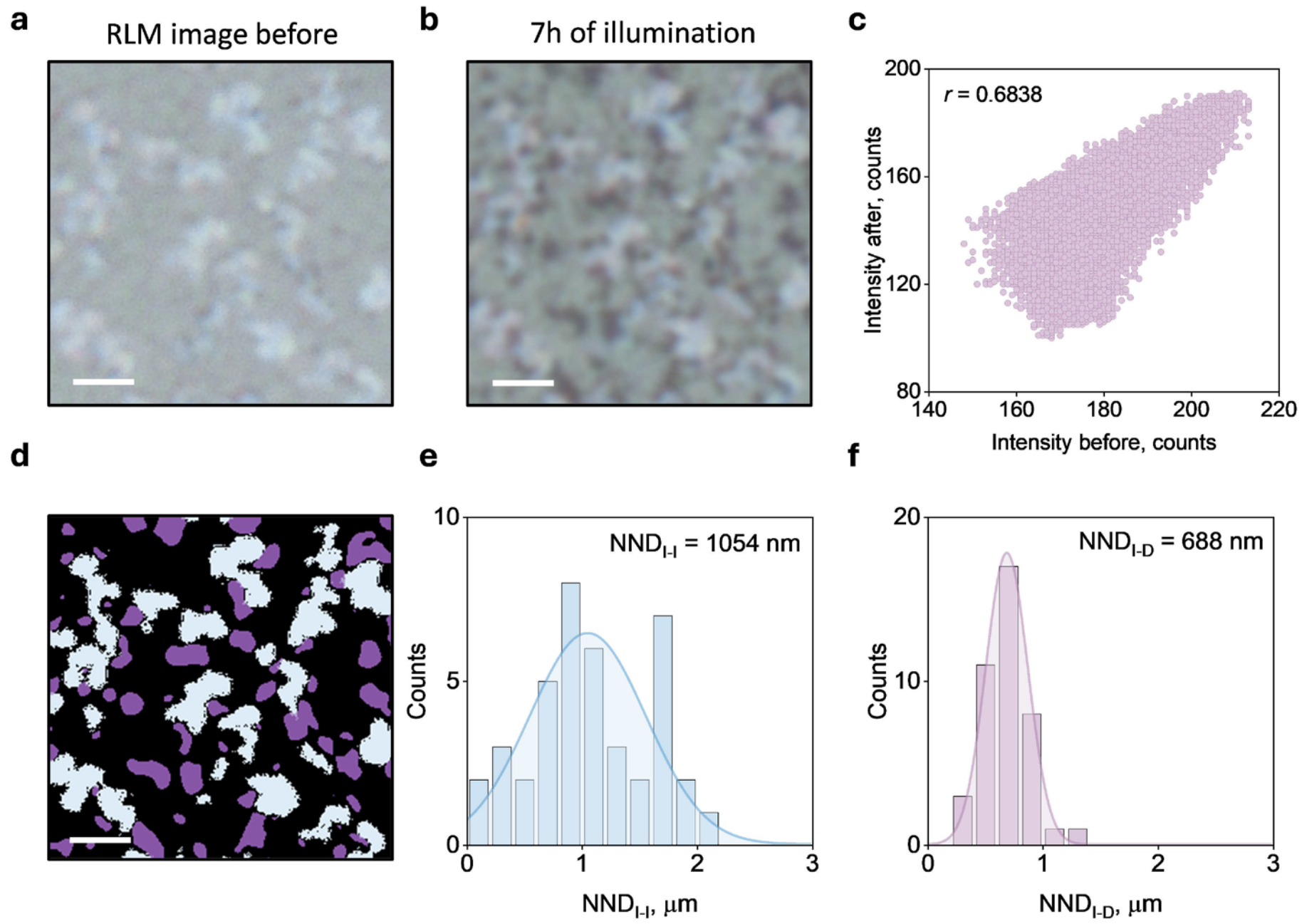

**Supplementary Fig. 22. Optical monitoring of degradation following prolonged glass-side illumination.** (a) RLM image of perovskite film before prolonged photo-stress. (b) RLM image of the same region following 7 hours of photo-stress with 3 Sun equivalent intensity. (c) RLM image intensity correlations before-after degradation for (a, b) with *r*. (d) Impurity-degradation map for (a, b). (e) $NND_{I-I}$ and (f) $NND_{I-D}$ distributions for (a). Scale bars are 2 μm.

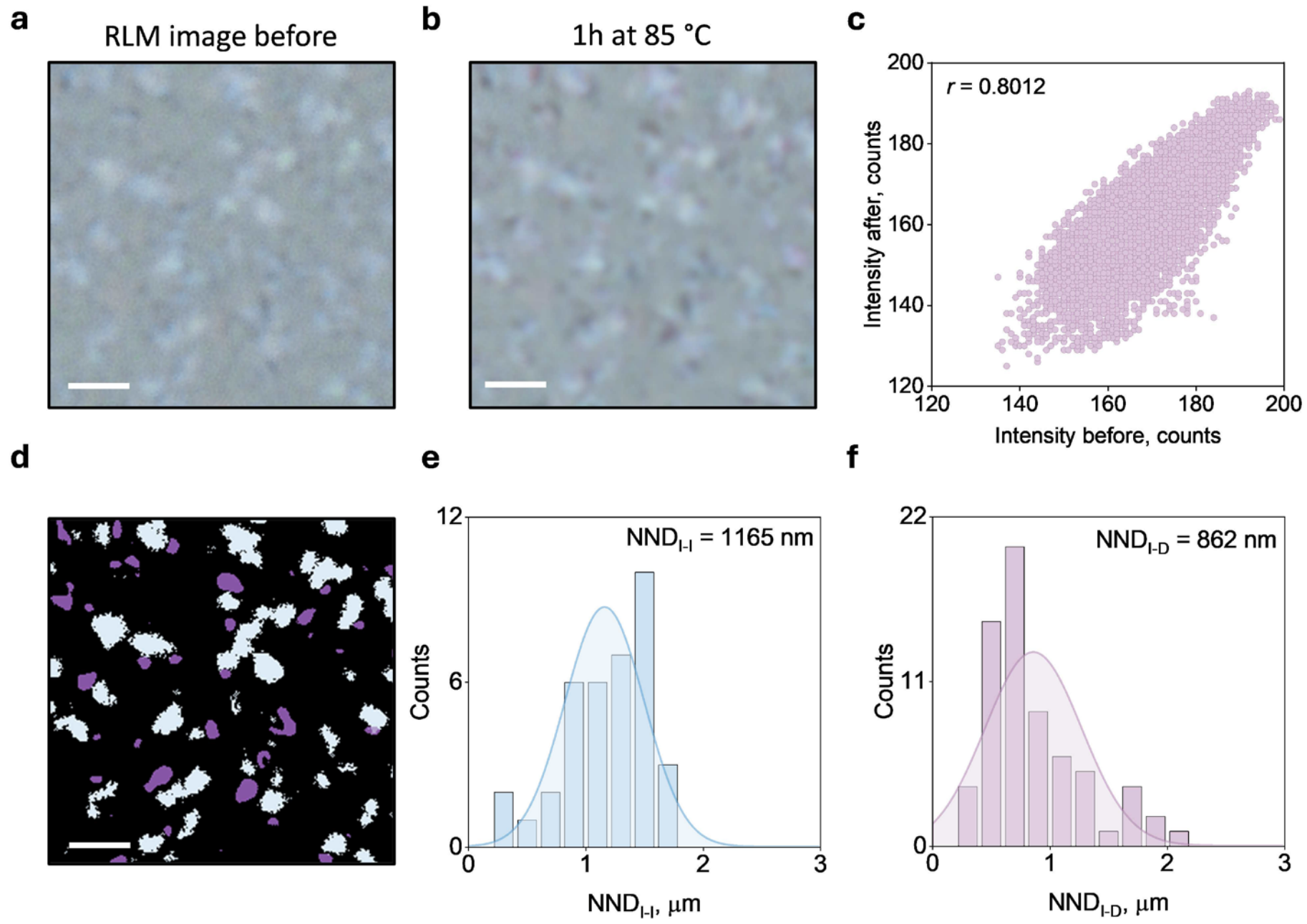

**Supplementary Fig. 23. Optical monitoring of degradation following thermal stress.** (a) RLM image of perovskite film before thermal stress. (b) RLM image of the same region following 1 hour of thermal stress at 85 °C. (c) RLM image intensity correlations before-after degradation for (a, b) with *r*. (d) Impurity-degradation map for (a, b). (e) $NND_{I-I}$ and (f) $NND_{I-D}$ distributions for (a). Scale bars are 2 μm.

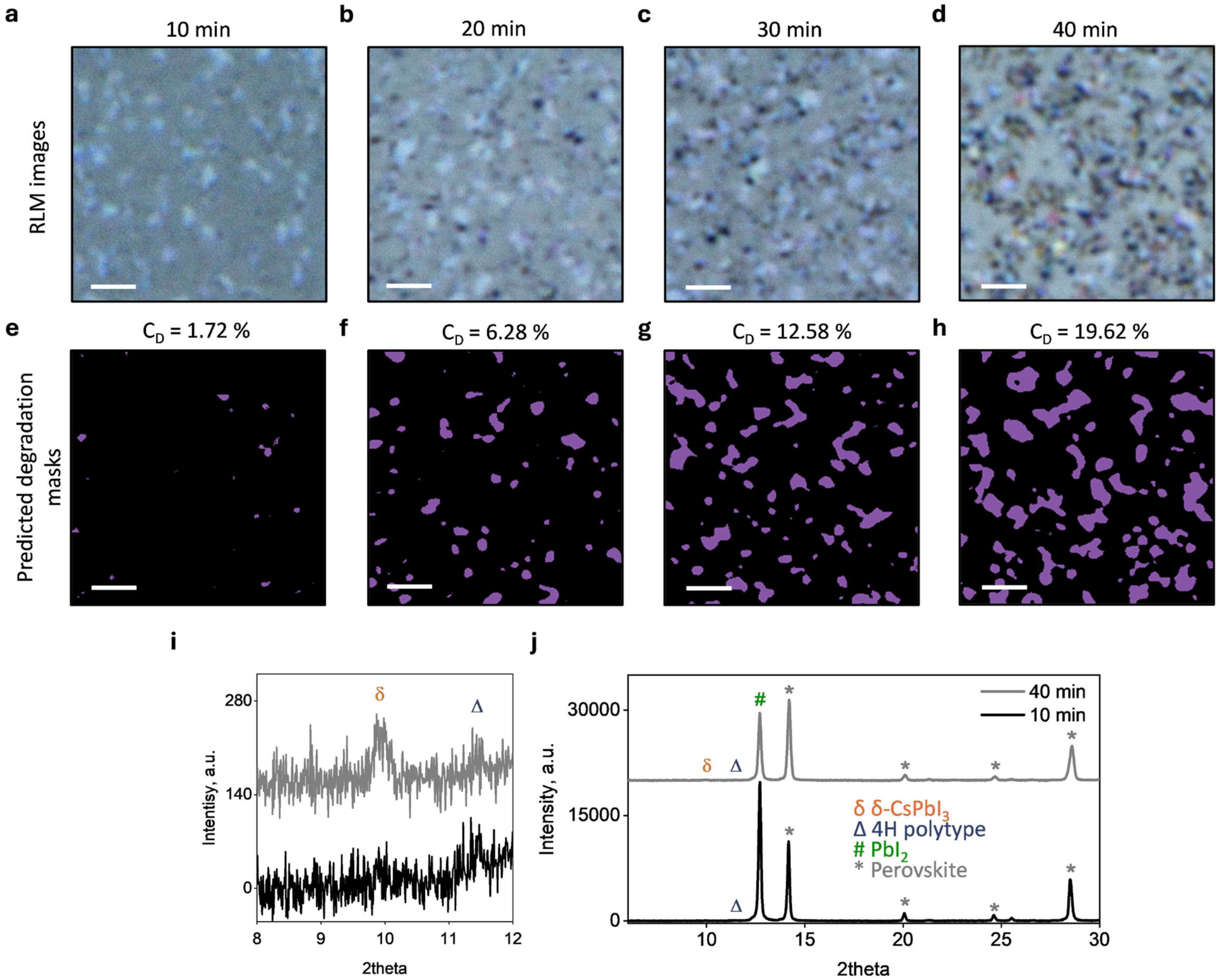

**Supplementary Fig. 24. Impact of annealing duration on local variation in optical contrast.** (a-d) RLM images of blade-coated perovskite films with varied annealing durations (10 min, 20 min, 30 min and 40 min, respectively). (e-f) Predicted degradation masks for (a-d). $C_D$ indicates total degradation coverage calculated for each mask. (i) Magnified XRD for 10 min and 40 min films, showing appearance of a δ-$CsPbI_3$ peak in the 40 min film, consistent with progressive increase in local darkening of optical contrast in RLM images. Scale bars are 2 µm.

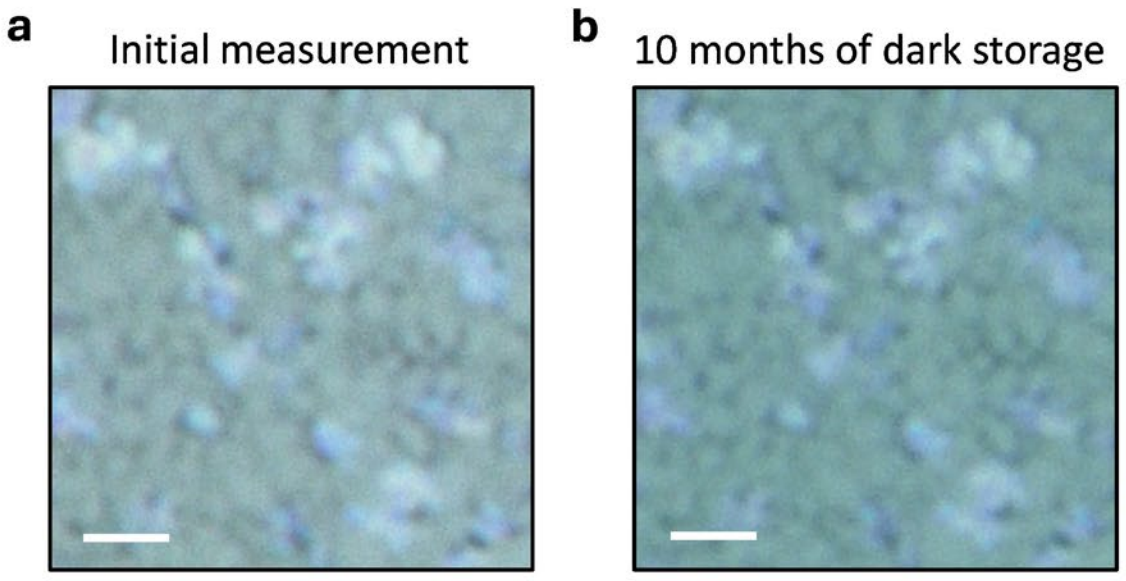

**Supplementary Fig. 25. Optical monitoring of perovskite films following dark storage.** (a) RLM image of a blade-coated perovskite film. (b) RLM image of the same region following 10 months of storage in dark $N_2$ environment, showing no changes to the film morphology and RLM contrast. The overall image color in (b) was adjusted to compensate for different light sources and cameras used for imaging. Scale bars are 2 μm.

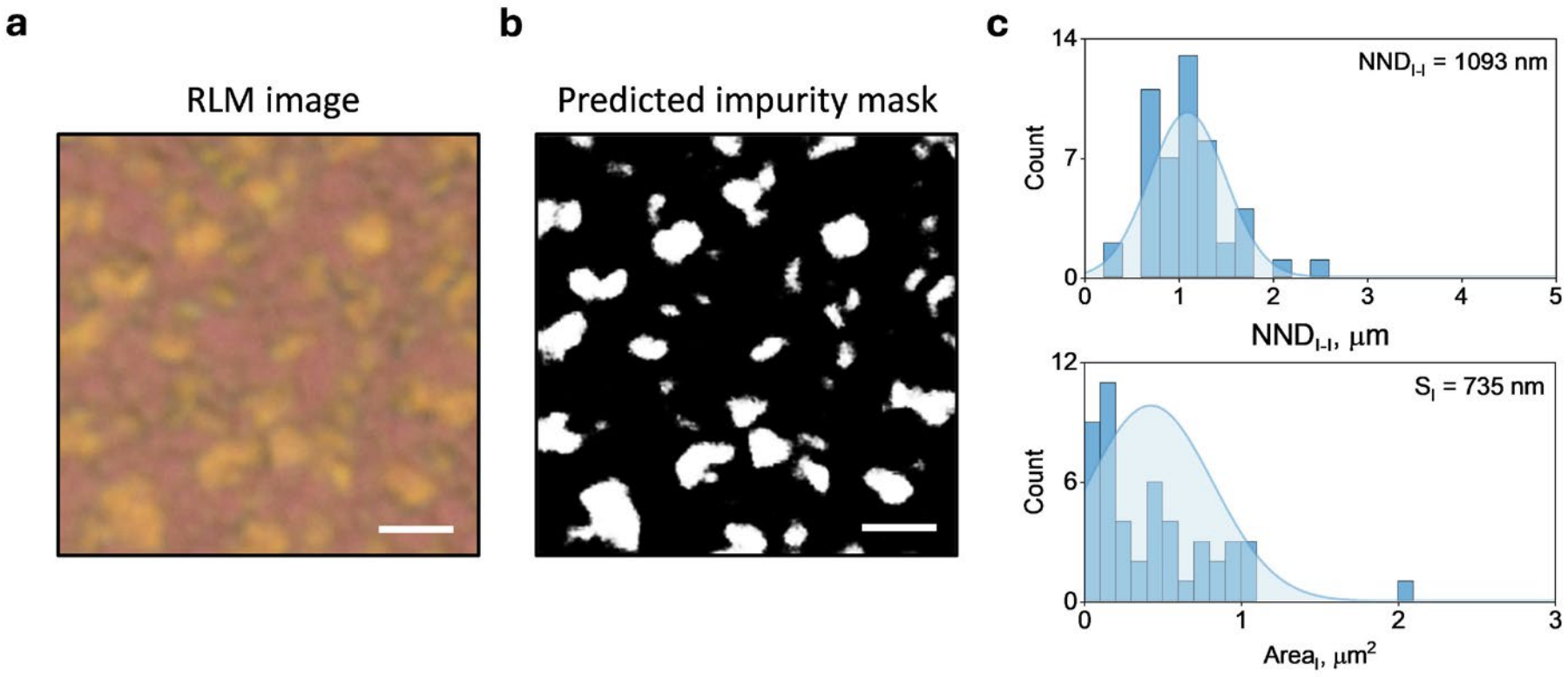

**Supplementary Fig. 26. RLM imaging of a semi-transparent perovskite solar cell**. (a) RLM image of a semi-transparent single-junction perovskite solar cell with glass/ITO/SAM/perovskite/$C_{60}$/$SnO_2$/IZO architecture. (b) Predicted impurity mask for (a). (c) $NND_{I\text{-}I}$ and $S_I$ distributions for (a). Scale bars are 2 µm.

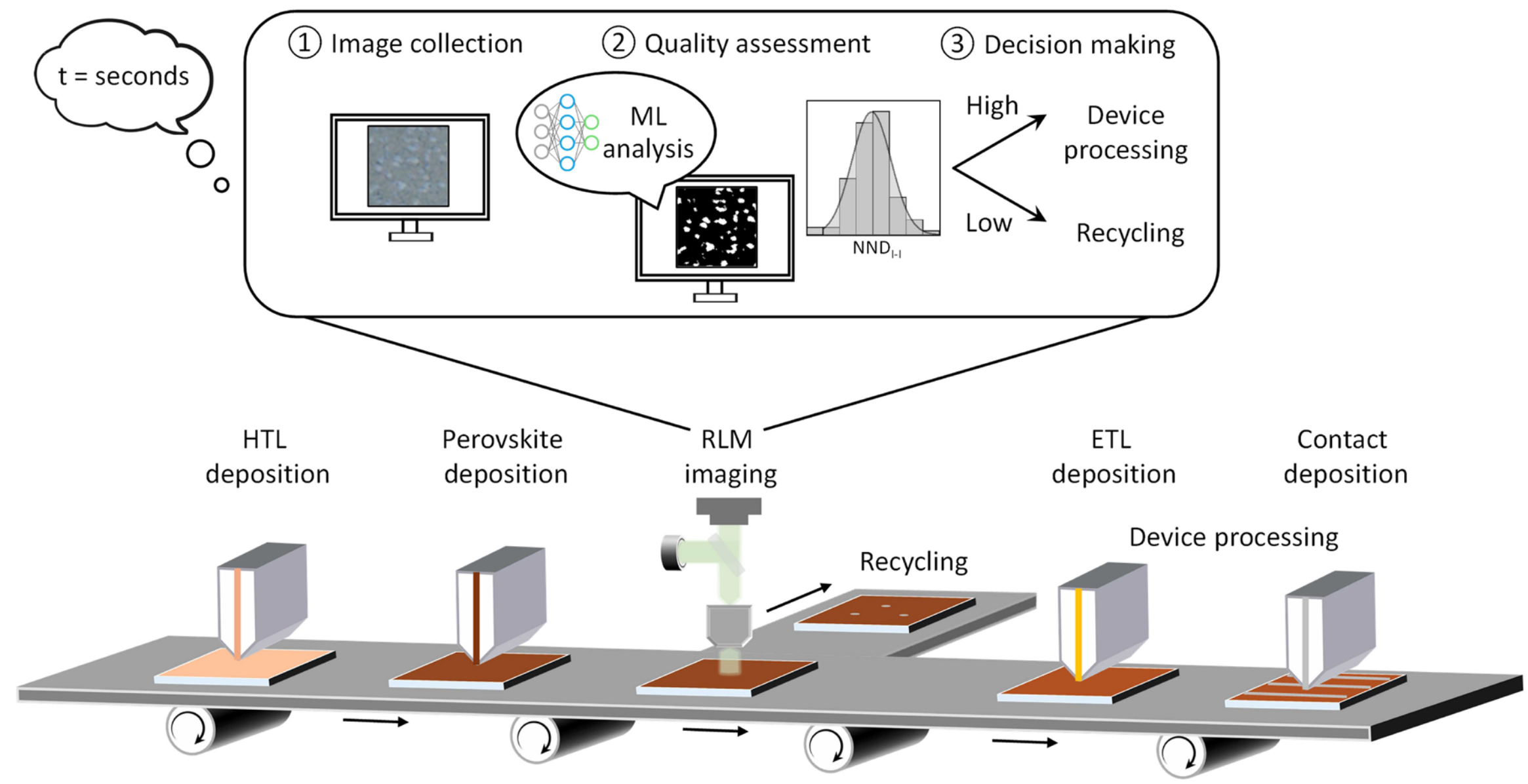

**Supplementary Fig. 27. Schematic of early-stage in-line pre-screening of perovskite films with high-throughput RLM.** RLM imaging is introduced to device-relevant stacks finished with a perovskite layer after deposition of the hole-transport layer (HTL). Rapid imaging and analysis with total time (t) on the order of seconds enables fast decision-making where high-quality films with low impurity density ($NND_{I-I}$) are selected for further device processing (e.g., deposition of the electron-transport layer (ETL) and contacts), while low-quality films with high impurity density undergo recycling (removal of perovskite layer and subsequent reuse of transparent conductive oxide substrates).

## 3. Supplementary References